\pdfoutput=1
\documentclass[12pt,onecolumn]{IEEEtran}

\usepackage[utf8]{inputenc}
\usepackage{graphicx}
\usepackage[T1]{fontenc}
\usepackage{url}
\usepackage{ifthen}
\usepackage[cmex10]{amsmath} 
\usepackage{amsmath}
\usepackage{setspace,color}
\usepackage{amssymb}
\usepackage{multicol}

\makeatletter
\newtheorem{definition}{Definition}
\newtheorem{theorem}{Theorem}
\newtheorem{proposition}{Proposition}
\newtheorem{lemma}{Lemma}
\newtheorem{example}{Example}
\newtheorem{corollary}{Corollary}
\newtheorem{remark}{Remark}
\makeatother

\def\a{{\alpha}}

\newcommand{\defined}{\triangleq}

\newcommand{\graph}{\set{G}}
\newcommand{\nodes}{\set{V}}
\newcommand{\edges}{\set{E}}
\newcommand{\N}{\set{N}}

\newcommand{\X}{\set{X}}
\newcommand{\A}{\set{A}}

\newcommand{\Z}{\set{Z}}

\newcommand{\hb}[1]{
{h_{b}\left( #1 \right)}
}

\begin{document}

\newcommand{\seq}[3]{{#1_{#2}, \ldots, #1_{#3}}}
\newcommand{\newsym}[1]{#1}
\newcommand{\newnot}[1]{#1}
\def\defined{\triangleq}
\def\sessions{{\cal S}}

\title{Capacity Bounds for Networks with Correlated Sources and Characterisation of Distributions by Entropies}
\author{\IEEEauthorblockN{ Satyajit Thakor~\IEEEmembership{Member, IEEE},
    Terence Chan~\IEEEmembership{Member, IEEE} and~Alex Grant~\IEEEmembership{Senior Member,~IEEE}} \thanks{S. Thakor is with School of Computing and Electrical Engineering, Indian Institute of Technology Mandi. T. Chan is with the Institute for Telecommunications Research, University of South Australia and A. Grant is with Myriota Pty Ltd. The material in this paper was presented in part at the Australian Communication Theory Workshop, Melbourne, Australia, 2011 \cite{ThaChaGra11}, Information Theory Workshop, Seville, Spain, 2013 \cite{ThaChaGra13} and accepted in International Symposium on Information Theory and its Applications, California, USA, 2016 \cite{ThaChaGra16}. 
    T. Chan is supported in part by the
    Australian Research Council under Discovery Projects DP150103658.}}
    
\maketitle

\begin{abstract}
Characterising the capacity region for a network can be extremely difficult. Even with independent sources, determining the capacity region can be as hard as the open problem of characterising all information inequalities. The majority of computable outer bounds in the literature are relaxations of the Linear Programming bound which involves entropy functions of random variables related to the sources and link messages.  When sources are not independent, the problem is even more complicated.  Extension of Linear Programming bounds to networks with correlated sources is largely open. Source dependence is usually specified via a joint probability distribution, and one of the main challenges in extending linear program bounds is the difficulty (or impossibility) of characterising arbitrary dependencies via entropy functions. This paper tackles the problem by answering the question of how well entropy functions can characterise correlation among sources.  We show that by using carefully chosen auxiliary random variables, the characterisation can be fairly ``accurate''. Using such auxiliary random variables we also give implicit and explicit outer bounds on the capacity of networks with correlated sources. The characterisation of correlation or joint distribution via Shannon entropy functions is also applicable to other information measures such as R\'{e}nyi entropy and Tsallis entropy.
\end{abstract}


\newcommand{\dist}[1]{{\Omega}_{#1}}
\newcommand{\setindex}[1]{{ {\mathcal P}_{[2,#1]} }}

\def\lr{{\langle}}
\def\rr{{\rangle}}
\def\graph{{\mathcal G}}
\def\nodes{{\mathcal V}}
\def\edges{{\mathcal E}}
\newcommand\eht[2]{{#1 \to #2}}
\def\A{{\mathcal A}}

\section{Introduction}

The fundamental question in network coding is to determine the
required link capacities to transmit the sources to the
sinks. Characterising the network coding capacity region is extremely
hard~\cite{ChaGra08}. Despite its importance, the maximal gain that can be obtained by network coding is still largely unknown, except in a few scenarios \cite{AhlCai00,YeuZha99}. One example is the single-source scenario where the capacity region is characterised by
the max-flow bound \cite{AhlCai00} (see also \cite[Chapter 18]{Yeu08}) and linear network codes maximise throughput \cite{LiYeu03}. However, when it involves more than one source, the problem can become quite difficult.

The problem becomes even more complex when the sources are correlated. When the sources are independent, the
capacity region depends only on the source entropy rates. However,
when the sources are dependent, the capacity region depends on the
detailed structure of the joint source distribution. In the classical literature, the problem of communicating correlated sources is called distributed source compression \cite{SleWol73,WynZiv76}. For networks, the distributed source compression problem is a feasibility problem: given a network with edge capacity constraints and the joint probability distribution of correlated sources available at certain nodes, is it feasible to communicate the correlated sources to demanding nodes?

A relevant important problem is of separation of distributed source coding and network coding \cite{RamJaiChoEff06}. Specifically,  distributed source coding and network coding are separable if and only if optimality is not sacrificed  by separately designing source and network codes. It has been shown in \cite{RamJaiChoEff06} that the separation holds for two-source two-sink networks however it has been shown by examples that that the separation fails for two-source three-sink and three-source two-sink networks. 

In \cite{Han11}\footnote{The results were generalised for networks with noisy channels. However, in this paper we are mainly concerned with networks with error-free channels.}, Han gave a necessary and sufficient condition for the set of achievable rates when each sink requires all the sources (see also \cite{RamJaiChoEff06} for noiseless channel network model). 
This result includes the necessary and sufficient condition \cite{Han80}, \cite{BarSer06} for networks in which every source is demanded by single sink as a special case.
Until recently there did not even exist in the literature a nontrivial necessary condition for reliable transmission of correlated sources in general multicast networks. In \cite{ThaGraCha16a}, we made the first attempt to address this problem by characterising a graph based bound, called the ``functional dependence bound'',\footnote{The functional dependence bound was initially characterised for networks with independent sources in \cite{ThaGraCha09}.} for networks with correlated sources with arbitrary sink demands. The functional dependence bound \cite{ThaGraCha16a,ThaGraCha09} is tighter than the cut-set bound \cite{CovTho06}.

Following \cite{Yeu02},  we develop a linear programming outer bound for dependent sources (see Theorem \ref{thm:R_Out} in this paper).  This bound is specified by a set of information inequalities and equalities, and source dependence is represented by the entropy function 
\begin{align}\label{1}
h(\alpha) \triangleq H(Y_{s}^{n}, s\in\alpha), \alpha \subseteq \sessions
\end{align}
where $\sessions \triangleq \{1,\ldots,|\sessions|\}$ is an index set for the sources and 
\[
\{(Y_{s}^{n},
s\in \sessions), n = 1, 2,\ldots\}
\] 
are independent and
identically distributed copies of the $|\sessions|$ dependent sources. Thus
each $(Y_{s}^{n}, s\in \sessions)$ has the same joint distribution
as the sources, but are independent across different $n$. However, within the same ``time'' instance $n$, the random variables $(Y_{s}^{n}, s \in \sessions)$ may be correlated among different $s$.

The concurrent work \cite{GohYanJagg13} (see also \cite{GohYanJagg11}) focused on improving the cut-set bound for networks with correlated sources. For this a so called ``uncertainty region'' was proposed and characterised. 
For two source case, with random variables $Y_{s}^{(N)} \defined (Y_s^n, n = 1 \ldots N), s \in \{1,2\}$,  the uncertainty region is the closure of the set of all four-dimensional vectors
$$\bigg[\frac{H(K)}{N}, \frac{H(K|Y_1^{(N)})}{N},\frac{H(K|Y_2^{(N)})}{N},\frac{H(K|Y_1^{(N)},Y_2^{(N)})}{N} \bigg]$$
where  the random variable $K$ may be interpreted as ``almost'' common information when the last three quantities in the vector are very small. Independently, we also constructed in \cite{ThaChaGra11} auxiliary random variables which are almost common information to characterise tighter outer bounds on network capacity (see Section \ref{sec:Using common information} of this paper). The uncertainty region was used to improve cut-set based converse theorems for network capacity. See Section \ref{sec:Extensions} for a comparison of the work in \cite{GohYanJagg13} and the work in this paper.

The simple formulation in \eqref{1} fails to properly characterise source dependence.
We also note that the capacity regions (or best known achievable
regions) for many classic multiterminal problems are also expressed as
optimisations of linear combinations of joint entropies, subject to
linear constraints (e.g., markov constraints) on joint entropies. If it
were not for the specified joint distributions on the
sources/side-information etc., typically present in such problems,
numerical solution would be achieved by a linear program. Again, if it
were possible to somehow accurately capture the dependence of random variables
using entropies, it would lead to a convenient computational approach. A natural question arises: \emph{How accurately can arbitrary dependencies be specified via entropies alone?} We show that by using auxiliary random variables, entropies can in fact be sufficient.

\subsubsection*{Organization}
In Section \ref{sec:Motivating Applications} we present bounds on the capacity of networks with correlated sources. In particular, we characterise outer bounds using geometrical approach (referred as geometric bounds) in Section \ref{sec:Network Coding Capacity Outer Bounds}. Section \ref{sec:Tightening the bounds by using auxiliary variables} demonstrates that these bounds are not tight and can be tightened by introducing new auxiliary random variables which more accurately describe correlation between the source random variables. We also give a general framework (Definition \ref{def:Improved Outer Bound} and Theorem \ref{thmouterbd2}) for improving outer bounds with introduction of auxiliary random variables. Section \ref{sec:Implicit and Explicit Outer Bounds on Network Capacity} presents an implicit as well as an explicitly computable bound using the partition auxiliary random variables describing the source correlation in Section \ref{sec:Construction of Auxiliary Random Variables}. In Section \ref{sec:Two Constructions of Auxiliary Variables}, we present two approaches to construct auxiliary random variables to tighten the outer bounds. The constructions via these two approaches are direct generalisations of the auxiliary random variables designed for the example network in Section \ref{sec:Tightening the bounds by using auxiliary variables}. In Section \ref{sec:DistributionViaEntropy}, we deal with the more general problem of characterising probability distribution using entropy functions. Specifically, in Section \ref{sec:Construction of Auxiliary Random Variables} we give a characterisation of distributions via partition auxiliary random variables for scalar random variables and for vector random variables in Theorems \ref{thm1} and \ref{thm2} respectively. In Section \ref{sec:Extensions}, we briefly describe extension of our work to ``vector-block characterisation'' and to other information measures such as R\'{e}nyi entropy and Tsallis entropy. 


\section{Capacity Outer Bounds}\label{sec:Motivating Applications}
In this section, we focus on characterisation of network coding capacity outer bounds for networks with correlated sources.  Let the directed acyclic graph $\graph = (\nodes, \edges)$ serve as a simplified model of a communication network with error-free point-to-point communication links. Edges $e\in\edges$ have capacity $C_e>0$. 
Let $\{(Y_{s}^{n},
s\in \sessions), n = 1, 2, \ldots,   \}$  be the set for the correlated  sources. Here,  each  source is a stream of  identically distributed  source symbols. For each $n$, the sources symbols 
$(Y_{s}^{n}, s\in \sessions)$
are assumed to be correlated (with the same joint distribution), 
but are independent across different $n$. For simplicity, the superscript
$n$ will often be dropped.

The locations of the sources are identified by the mapping 
\begin{equation*}
a : \sessions \mapsto  \boldsymbol{\mathcal P}(\mathcal V).
\end{equation*}
(a source may be available at multiple nodes) and each source can be demanded by more than one sink nodes, characterised by the mapping
 \begin{equation*}
b : \sessions \mapsto \boldsymbol{\mathcal P}(\mathcal V).
\end{equation*}
Here, $\boldsymbol{\mathcal P}(\mathcal V)$ is the collection of all subsets of $\mathcal V$. For all $s$ assume that $a(s)\cap b(s)=\emptyset$. Each edge $e\in\edges$ in the network carries a random variable $U_e$  which corresponds to the message (or stream of messages) transmitted on that particular link. Let $e=(u,v)$ and $e'=(u',v')$. Then we will use the notation 
$e' \rightarrow e $ to denote the condition that the head of $e'$ and the tail of $e$ are the same (i.e., to denote that $v'=u$). Similarly, we will use $s \rightarrow e$ to denote that $u \in a(s)$, and  $e \rightarrow w$ to denote that $v=w$.  Using our notations, the message $U_{e}$ transmitted on link $e$ must be a function of all the sources $s$ such that $s \rightarrow e$ and transmitted messages on $e'$ where   $e' \rightarrow e $.

\begin{definition}[Network code]\label{def:networkcode}
A network code $\phi_{\graph}^{(N)}$ (over a block of $N$ symbols) for a given network $\graph=(\nodes,\mathcal E)$ is described by a set of local encoding functions 
\begin{align*} 
 \phi_{e}^{(N)} &: \prod_{s \in \mathcal S : s \rightarrow e} \mathcal Y_{s}^{(N)} \times \prod_{f \in \mathcal E: \eht{f}{e}} \mathcal U_{f}^{(N)}
 \longmapsto \mathcal U_{e}^{(N)}  
\end{align*}
for  $e \in \mathcal E$, and decoding functions
\begin{align*} 
 \phi_{u}^{(N)} &: \prod_{s' \in \mathcal S : u \in a(s')} \mathcal Y_{s'}^{(N)} \times \prod_{f \in \mathcal E: \eht{f}{u}} \mathcal U_{f}^{(N)}
 \longmapsto \mathcal Y_{s}^{(N)} 
\end{align*}
for $ u \in b(s)$ and $s \in \mathcal S$.

Here, the alphabets of the block of source random variables $Y_{s}^{(N)} = (Y_{s}^{n}, n=1, \ldots, N)$ is $\mathcal Y_{s}^{(N)}$ and  $\mathcal U_{e}^{(N)}$ is the alphabet set for the message being transmitted on link $e$. The function $\phi_{e}^{(N)}$ determines how the transmitted message $U_{e}^{(N)}$ will be encoded and the function $\phi_{u}^{(N)}$ dictates how the sources should be regenerated at the sink nodes. 
\end{definition}

\begin{definition}[Achievable rate tuple]
A link capacity tuple $\mathbf{{C}} = (C_{e}: e \in \mathcal E)$  is called \emph{achievable} if there exists a sequence of network codes $\phi_{\mathcal G}^{(N)}$ such that for every $e \in \mathcal E$ and every $s \in \mathcal S$
\begin{align}
\lim_{N \rightarrow \infty} \frac{\log |\mathcal U^{(N)}_{e}|}{N} &\leq C_{e}   \label{eq:sequenceofNC1}
\end{align}
and 
\begin{align}
\lim_{N \rightarrow \infty} \textrm{Pr}\{\phi_{u}^{(N)}(U^{(N)}_{f}: f \rightarrow u, Y^{(N)}_{s'}, u \in a(s')) \neq Y^{(N)}_s\} &= 0 \label{eq:sequenceofNC2}
\end{align}
for all $e\in\edges$,  $s\in\sessions$ and $u\in b(s)$.
\end{definition}

\begin{definition}[Achievable region]
The set of all achievable link capacity tuples will be denoted by $\mathcal R_{\text{cs}}.$\footnote{The subscript describes correlated source case.}
\end{definition}

\subsection{Network Coding Capacity Outer Bounds}\label{sec:Network Coding Capacity Outer Bounds}
Following \cite{Yeu02}, we first develop geometric\footnote{Geometric in a sense that the random variables in a network and the constraints define a region in an
Euclidean space and a bound is viewed as a region in the Euclidean space.} outer bounds for the achievable region.

\def\Z{{\cal Z}}
\def\R{{\cal R}_{\mathrm{cs}}}

\begin{definition}[Polymatroids]
Let $\Z = \sessions \cup \edges$. A function $h : {\mathcal P}(\Z) \mapsto {\mathbb R}$ is a polymatroid if 
\begin{align*}
h(\emptyset) & = 0 \\
h(\alpha) & \ge h(\beta) \ge 0, \quad \forall \beta\subseteq \alpha \subseteq \Z \\
h(\alpha) + h(\beta) & \ge h(\alpha \cap \beta) + h(\alpha \cup \beta), \quad \forall \alpha,\beta \subseteq \Z.
\end{align*}

\end{definition}

\begin{remark}
To simplify our notation, we will use $h(\alpha | \beta) $ to denote 
$h(\alpha \cup \beta) - h(\beta)$ for $\alpha, \beta \subseteq \Z$. 
\end{remark}

\begin{definition}\label{def:regionsCS}
Let $\Delta$ be a subset of polymatrods. Define 
$\R(\Delta)$ as the set of  all link capacity tuples $\mathbf{C} = (C_{e}: e \in \mathcal E)$ such that  there exists $h\in \Delta$ satisfying the following conditions
\begin{align}
h (\alpha)-H(Y_{s}, s\in \alpha)&=0 \label{eq:R_out1}\\
h ({e}| f \in \sessions\cup\edges \:: f \rightarrow  {e}  )  &= 0\\
h (s| s' \in \sessions  : u \in a(s'), f \in\edges : f \rightarrow u   ) &=0 \\
h ({e}) &\leq C_{e}  \label{eq:R_out5} 
\end{align}
for all $\alpha\subseteq \sessions$, $s\in\mathcal S$, $u\in b(s)$ and $e\in \mathcal E$.
\end{definition}

Taking $\Delta$ as $\overline{\Gamma^*}$ and $\Gamma$ in Definition \ref{def:regionsCS} gives us regions $\mathcal R_{\mathrm{cs}}(\overline{\Gamma^*})$ and $\mathcal R_{\mathrm{cs}}(\Gamma)$ respectively.

\begin{theorem}[Outer bound]\label{thm:R_Out}
Let $\Gamma$ be the set of all polymatroids and $\overline{\Gamma^*}$ be the set of all almost entropic functions. Then 
\begin{equation}\label{eq:boundsCS}
\mathcal R_{\mathrm{cs}} \subseteq \mathcal R_{\mathrm{cs}}(\overline{\Gamma^*}) \subseteq \mathcal R_{\mathrm{cs}}(\Gamma).
\end{equation}
\end{theorem}
\begin{IEEEproof}
Let  $\textbf{\textrm{C}} = (C_{e}: e \in \mathcal E)$ be an achievable link capacity tuples.
By definition, there exists a sequence of network codes $\phi_{\mathcal G}^{(N)}$ satisfying \eqref{eq:sequenceofNC1}-\eqref{eq:sequenceofNC2}. Hence for any $\epsilon  >0$ and any sufficiently large $N$ the network code $\phi_{\mathcal G}^{(N)}$ satisfies
\begin{align}
\frac{H(U^{(N)}_{e})}{N} \le \frac{\log |\mathcal U^{(N)}_{e}|}{N}  & \leq C_{e} + \epsilon   \label{eq:achievableC1}
\end{align}
and 
\begin{align*}
  \textrm{Pr}\{\phi_{u}^{(N)}(U^{(N)}_{f}, f \rightarrow u, Y^{(N)}_{s'}, u \in a(s')) \neq Y^{(N)}_s\} \leq \epsilon 
\end{align*}
for all $e\in\edges$,  $s\in\sessions$ and $u\in b(s)$.
 
By the definition of a network code in Definition \ref{def:networkcode}, it is clear that for any $e\in \edges$
\begin{equation}
H(U^{(N)}_{e}| Y^{(N)}_{s}, u \rightarrow e, u \in a(s), U^{(N)}_{f}: f \rightarrow e ) = 0.
\end{equation}

On the other hand, for any $s \in \sessions$ and $u\in b(s)$,  Fano's inequality implies that 
 $$H(Y^{(N)}_{s} | U^{(N)}_{f}, f \rightarrow u, Y^{(N)}_{s'}, u \in a(s')) \leq 1 + \epsilon  \log  |\mathcal Y^{(N)}_s|.$$


It is easy to choose  $\varrho (N,\epsilon)$ such that for any $\epsilon > 0$, 
\begin{align*}
\lim_{\epsilon \to 0} \lim_{N \to \infty}\varrho (N,\epsilon) = 0 
\end{align*}
and  
\begin{align*}
 H(Y^{(N)}_{s} | U^{(N)}_{f}, f \rightarrow u, Y^{(N)}_{s'}, u \in a(s'))   \le N \varrho (N,\epsilon).
\end{align*}

Let $ h^{(N,\epsilon)}$ be obtained by multiplying the entropy function of 
$$(Y^{(N)}_s :s \in \mathcal S, U^{(N)}_{e}:e \in \mathcal E)$$  
with the factor $1/N$. 
In other words, for any $\alpha\subseteq \sessions$ and $\beta\subseteq \edges$
\begin{align*}
h^{(N,\epsilon)}(\alpha,\beta) = \frac{1}{N} H(Y^{(N)}_s :s \in \alpha, U^{(N)}_{e}:e \in \beta).
\end{align*}
Then, for every $e \in \mathcal E$ and  $s \in \mathcal S$
\begin{align*}
h^{(N,\epsilon)}( \alpha )- H(Y_s, s\in \alpha ) & = 0\\
h^{(N,\epsilon)} ({e}|  f \in \sessions\cup\edges \:: f \rightarrow  {e}  \}) &= 0 \\
h^{(N,\epsilon)}(s | s' \in \sessions  : u \in a(s'), f \in\edges : f \rightarrow u  ) &\leq   \varrho (N,\epsilon)\\
h^{(N,\epsilon)} ({e})  &\leq     C_{e} + \epsilon
\end{align*}
for all $\alpha\subseteq \sessions$, $s\in\mathcal S$, $u\in b(s)$ and $e\in \mathcal E$.

Finally, let 
\begin{align}\label{eq25}
h = \lim_{\epsilon \to 0} \lim_{N \to \infty} h^{(N,\epsilon)}.
\end{align} 
It can be easily proved that $h$ is almost entropic\footnote{Strictly speaking, the limit may not exist. However, one can always pick a convergent subsequence. Therefore, for notation simplicity, we will simply define $h$ as the limit.},  and will satisfy all the conditions \eqref{eq:R_out1}--\eqref{eq:R_out5}.
The theorem is thus proved.
\end{IEEEproof}

If we examine the bound in Theorem \ref{thm:R_Out}, the correlation of the sources is captured by the relation
equality \eqref{eq:R_out1}. However, these entropic relations are not sufficient to capture precisely how the sources are correlated. As a result, it is possible  that the outer bounds $\mathcal R_{\text{cs}}(\overline{\Gamma^*})$ and $\mathcal R_{\text{cs}}(\Gamma)$  are simply not tight. 

In the next section, we illustrate how to tighten the bounds by deriving additional entropic relations to better capture the correlation among sources.
%
\subsection{Tightening the Bounds using Auxiliary Variables}\label{sec:Tightening the bounds by using auxiliary variables}
 
\begin{figure}[htbp]
\centering
  \includegraphics[scale=0.35]{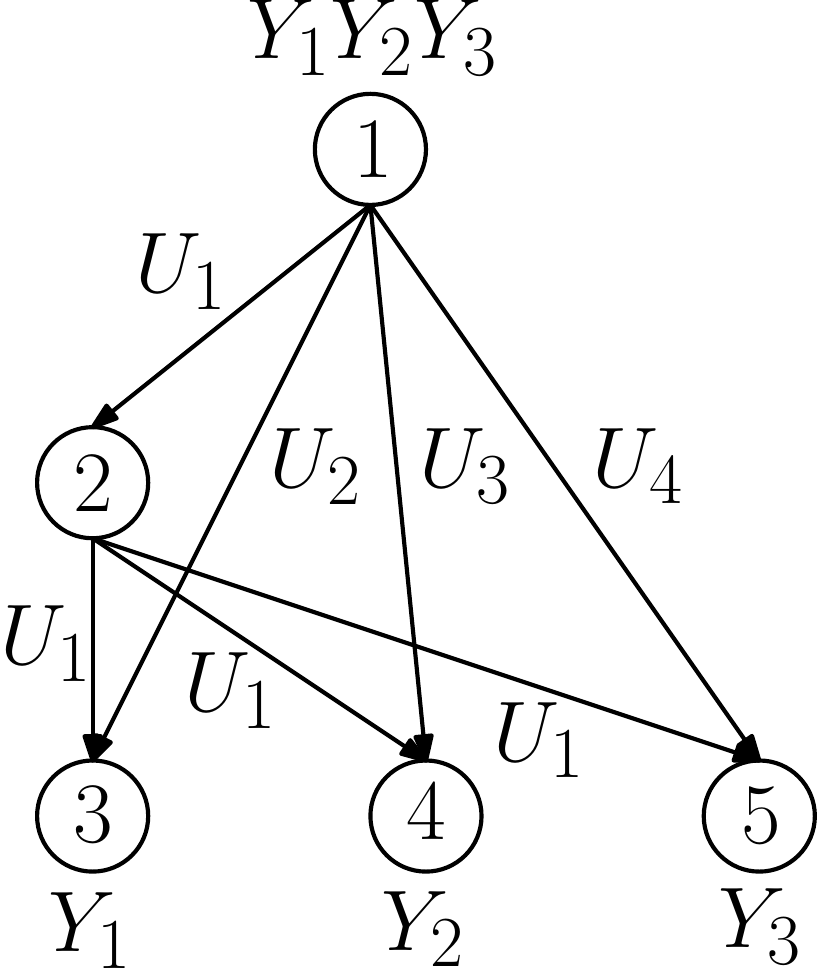}
  \caption{A network example.}\label{fig:net-cs}
\end{figure}

In Figure \ref{fig:net-cs}, three correlated sources $Y_1, Y_2, Y_3$ are available at node 1 and are demanded at nodes $3,4,5$ respectively. The edges from node $2$ to nodes $3,4,5$ have sufficient capacity to carry the random
variable $U_1$ available at node 2. The correlated sources $Y_1, Y_2, Y_3$ are defined as follows.
\begin{align*}
Y_1 = (b_0,b_1)\\
Y_2 = (b_0,b_2)\\
Y_3 = (b_1,b_2)
\end{align*}
where $b_0,b_1,b_2$ are independent, uniform binary random variables.

\begin{lemma}
For the network coding problem described above,  the outer bound 
$\mathcal R_{\text{cs}}(\Gamma)$  contains  all link capacity tuples $\textbf{\textrm{C}} =(C_i, i=1,...,4)$ such that there exists $ h \in \Gamma$  satisfying the following constraints.
\begin{align}
 h (s_{i})&=2, \forall i=1,2,3 \label{eq:ex-cs1}\\
 h (s_{i},s_{'j}) &=3, \forall i\neq j \\
 h (e_{i}|s_{1} , s_{2}, s_{3})&=0, i=1,2,3,4\\
 h (s_1|e_{1},e_{2})&=0\\
 h (s_2|e_1,e_3)&=0\\
 h (s_3|e_1,e_4)&=0\\
 h (e_{i})  &\leq C_i, i=1,...,4 \label{eq:ex-cs8}
\end{align}
\end{lemma}
\begin{IEEEproof}
A direct verification.
\end{IEEEproof}

Let $h$ be the entropy function of the following random variables:
\begin{align*}
Y_{s_{1}} &= (b_0,b_1),& U_{e_{1}} & =b_{0}\\
Y_{s_{2}} &= (b_0,b_2) & U_{e_{2}} & = b_{1}\\
Y_{s_{3}} &= (b_0,b_1 \oplus b_2) & U_{e_{3}} & = b_{2} \\
& & U_{e_{4}} & = b_{1} \oplus b_{2}.
\end{align*}
It can be verified easily that $h$ satisfies \eqref{eq:ex-cs1}-\eqref{eq:ex-cs8}. Using $h$, we prove that the link capacity tuple 
\begin{align}\label{eq38}
(C_i=1, i=1,...,4) \in \mathcal R_{\text{cs}}(\Gamma).
\end{align}

\def\bfh{{h}}

In the following, we will describe how to tighten the previous linear programming bound by better capturing the correlation among sources via additional entropic relations. We will then use the improved LP bound to show that the link capacity tuple in \eqref{eq38} is indeed not achievable.

In our first outer bound, the correlation among sources are simply characterised by the joint entropies of the source random variables. The idea behind our improved LP bound is by constructing and using auxiliary random variables.

\def\L{{\mathcal L}}

\begin{definition}\label{def:Improved Outer Bound}
Consider a set of correlated sources $Y_s, s\in {\mathcal S}$ with underlying probability distribution $P_{Y_{\mathcal S}}(\cdot)$. Construct   auxiliary random variables $K_{i},i\in \mathcal L$ by picking some  conditional probability distribution function $P_{K_{\mathcal L}|Y_{\mathcal S}}(\cdot)$. Let $\Delta^{*}$ be a subset of polymatrods over the set $\Z = \sessions \cup \edges \cup \L$. Define 
$\R^{*}(\Delta^*)$ as the set of  all link capacity tuples $\mathbf{C} = (C_{e}: e \in \mathcal E)$ such that  there exists $h\in \Delta^{*}$ satisfying the following conditions
\begin{align}
h (\alpha,\beta)-H(Y_{s},s\in \alpha,  K_{i}, i\in\beta )&=0 \label{eq:R_out*1}\\
h ({e}| f \in \sessions\cup\edges \:: f \rightarrow  {e}  )  &= 0\\
h (s| s' \in \sessions  : u \in a(s'), f \in\edges : f \rightarrow u   ) &=0 \\
h ({e}) &\leq C_{e}  \label{eq:R_out*5} 
\end{align}
for all $\alpha\subseteq \sessions$, $\beta\subseteq \L$, $s\in\mathcal S$, $u\in b(s)$ and $e\in \mathcal E$.
\end{definition}

\begin{theorem}[Improved Outer bounds]\label{thmouterbd2}
\begin{equation*}
\mathcal R_{\mathrm{cs}} \subseteq
\mathcal R^{*}_{\mathrm{cs}}(\overline{\Gamma^*}) \subseteq \mathcal R_{\mathrm{cs}}(\overline{\Gamma^*})\subseteq \mathcal R_{\mathrm{cs}}(\Gamma)
\end{equation*}
and 
\begin{equation*}
\mathcal R_{\mathrm{cs}} \subseteq \mathcal R^{*}_{\mathrm{cs}}(\Gamma) \subseteq \mathcal R_{\mathrm{cs}}(\Gamma).
\end{equation*}
\end{theorem}

\begin{IEEEproof}
The proof for the theorem is essentially the same as that in Theorem \ref{thm:R_Out} by  treating the auxiliary random variables $K_{i}, i\in \L$ as   virtual sources that are not available at and are not demanded by any nodes in the network.
\end{IEEEproof}

In the following, we will use the improved outer bound to show that the link capacity tuple \eqref{eq38} is not achievable.

\begin{lemma}\label{lemma:ImrovedLPnet-cs}
An improved LP bound $\mathcal R^{*}_{\text{cs}}(\Gamma)$ for the network in Figure \ref{fig:net-cs} is the set of all link capacity tuples $\textbf{\textrm{C}} =(C_e, e=1,...,4)$ such that there exists $ h \in \Gamma$  satisfying the following constraints.
\begin{align}
 h (s_{i})&=2, i=1,2,3  \label{eq:iLPs}\\
 h (s_i,s_j) &=3, i \neq j, i,j \in \{1,2,3\}\\
 h (k_i, i\in \alpha)&=|\alpha|, \alpha \subseteq \{0,1,2\}  \\
h({s_{1}|k_{0},k_{1}}) &=0 \\
h({s_{2}|k_{0},k_{2}}) &=0 \\
h({s_{3}|k_{1},k_{2}}) &=0 \\
h(k_0,k_1)&=h(s_1)\\
h(k_0,k_2)&=h(s_2)\\
h(k_1,k_2)&=h(s_3)\\
 h (e_{i}|s_1,s_2,s_3)&=0, i=1,2,3,4\\  
 h (s_1|e_{1},e_{2})&=0\\
 h (s_2|e_1,e_3)&=0\\
 h (s_3|e_1,e_4)&=0\\
 h (e_{i})  &\leq C_i, i=1,...,4.  \label{eq:iLPe}
\end{align}
\end{lemma}
\begin{IEEEproof}
The lemma follows from  Theorem \ref{thmouterbd2} by choosing 
$K_{1} , K_{2}, K_{3}$ as $b_{0}, b_{1}$ and $b_{3}$.
\end{IEEEproof}

Now, we will use the outer bound obtained in Lemma \ref{lemma:ImrovedLPnet-cs} to show that 
$
\textbf{\textrm{C}} = (C_i=1, i=1,...,4)  
$ 
is not achievable by showing that $\textbf{\textrm{C}} \not\in  \mathcal R^{*}_{\text{cs}}(\Gamma)$ defined in the above lemma.

Suppose to the contrary that $\textbf{\textrm{C}} \in \mathcal R^{*}_{\text{cs}}(\Gamma)$. Then by definition, there exists  a polymatroid $h$  satisfying \eqref{eq:iLPs}-\eqref{eq:iLPe}.
From those constraints, it is easy to prove that
\begin{align}
\bfh(e_{1}|k_0,k_1)&=0\nonumber\\
\bfh(e_{1}|k_0,k_2)&=0\nonumber\\
\bfh(e_{1}|k_1,k_2)&=0 \nonumber\\
\bfh(k_{0},k_{1}, k_{2}) & = \bfh(k_{0}) + \bfh(k_{1}) + \bfh(k_{1}). \label{eq:4}
\end{align}
As $\bfh(e_{1}|k_0,k_2) = 0$, it implies that
\begin{equation*}
I_{\bfh}(e_{1};b_{1}|b_0,b_2) \triangleq h(e_{1}| b_{0},b_{2}) - h(e_{1}| b_{0},b_{2},b_{1})= 0.
\end{equation*}

On the other hand, by \eqref{eq:4}, we have
\begin{equation*}
I_{\bfh}(b_{1} ; b_{2} |b_{0} ) = 0.
\end{equation*}
Therefore,
\begin{equation*}
I_{\bfh}(b_{1} ; b_{2}, e_{1} | b_{0} ) = 0
\end{equation*}
and consequently,
\begin{equation*}
I_{\bfh}(b_{1} ; e_{1} | b_{0} ) = 0.
\end{equation*}
Together with $\bfh(e_{1} | b_{0},b_{1}) = 0$, this implies
$\bfh(e_{1} | b_{0}) = 0$. Similarly, we can also prove
that
\begin{equation*}
\bfh(e_{1} | b_{1})  = \bfh(e_{1} | b_{2}) = 0.
\end{equation*}
Together with $\bfh(b_{1}b_{2}) = \bfh(b_{0}) + \bfh(b_{1})$, 
we can then prove that  $h(e_{1}) = 0$.

Finally,
$\bfh(s_{1}  | e_{1},e_{2}) = 0$ implies
\begin{align*}
2 = \bfh(s_{1}) & \le \bfh( e_{1},e_{2}) \\
& \le \bfh( e_{1}) + \bfh( e_{2}) \\
& = \bfh( e_{2})  \\
& \le 1.
\end{align*}
A contradiction occurs. Thus, we prove that 
the  link capacity tuple $(C_e=1, i=1,...,4) \not \in \mathcal R^{*}_{\text{cs}}(\Gamma)$ and hence is not achievable. This example shows that the bound 
$\mathcal R^{*}_{\text{cs}}$ in Lemma \ref{lemma:ImrovedLPnet-cs}  is in fact tighter.

\subsection{Implicit and Explicit Outer Bounds on Network Capacity}\label{sec:Implicit and Explicit Outer Bounds on Network Capacity}
The bounds $\mathcal R^{*}_{\mathrm{cs}}(\overline{\Gamma^*})$ and $\mathcal R^{*}_{\mathrm{cs}}(\Gamma)$ in Theorem \ref{thmouterbd2} are implicit in a sense that exact construction of auxiliary random variables is not given. 

Using the characterisation of distribution for vector random variable via entropy functions of partition random variables described in Section \ref{sec:DistributionViaEntropy}, we now describe an implicit outer bound $\mathcal R'_{\mathrm{cs}}(\overline{\Gamma^*})$ (implicit since $\overline{\Gamma^*}$ has only implicit characterization yet) and an explicit outer bound $\mathcal R'_{\mathrm{cs}}(\Gamma)$ on the capacity of networks with correlated sources as follows.
\begin{definition}\label{def:Improved Outer Bound-vector}
Consider a set of correlated sources $Y_s, s\in {\mathcal S}$ with underlying probability distribution $P_{Y_{\mathcal S}}(\cdot)$. From this distribution, construct binary partition random variables
$A_{\langle \alpha \rangle}, \langle \alpha \rangle \in \Omega$ from partitions of $\mathcal Y_{\mathcal S}$ as described in Theorem \ref{thm2}. 
Let $\mathcal R'_{\mathrm{cs}}(\overline{\Gamma^*})$ be the set of all link capacity tuples $\mathbf{{C}} = (C_{e}: e \in \mathcal E)$ such that there exists an almost entropic function $ h  \in \overline{\Gamma^*}$ for the set $\sessions \cup \edges \cup \Omega$  satisfying the constraints
\begin{align*}
h (\alpha, \beta)-H(Y_{s}, s \in \alpha, A_{i}, i \in \beta)&=0\\
h ({e}| f \in \sessions\cup\edges \:: f \rightarrow  {e}  )  &= 0\\
h (s| s' \in \sessions  : u \in a(s'), f \in\edges : f \rightarrow u   ) &=0 \\
h ({e}) &\leq C_{e} 
\end{align*}
for all $\alpha \subseteq \mathcal S, \beta \subseteq \Omega, s \in \sessions, u \in b(s)$ and $e\in \mathcal E$. Replacing $\overline{\Gamma^*}$ by $\Gamma$ in Definition \ref{def:Improved Outer Bound-vector} we obtain an \emph{explicitly computable outer bound} $\mathcal R'_{\mathrm{cs}}(\Gamma)$.
\end{definition}

Following is a corollary of Theorem \ref{thmouterbd2}.
\begin{corollary}
\begin{equation*}
\mathcal R_{\mathrm{cs}} \subseteq
\mathcal R'_{\mathrm{cs}}(\overline{\Gamma^*}) \subseteq \mathcal R_{\mathrm{cs}}(\overline{\Gamma^*})\subseteq \mathcal R_{\mathrm{cs}}(\Gamma)
\end{equation*}
and 
\begin{equation*}
\mathcal R_{\mathrm{cs}} \subseteq \mathcal R'_{\mathrm{cs}}(\Gamma) \subseteq \mathcal R_{\mathrm{cs}}(\Gamma).
\end{equation*}
\end{corollary}

\section{Two Constructions of Auxiliary Variables}\label{sec:Two Constructions of Auxiliary Variables}
In the previous section, the bound for the network coding region can be formulated as a linear programming problem, in which  the correlation among the sources are captured by some entropic relations. 
To better capture the correlation, we have proposed the use of auxiliary random variables to obtain more entropic relations. 
We also gave examples showing that how this can indeed tighten the bound. 

The question however is how one should choose or define the auxiliary random variables. In some cases like in the previous network example, the choice can be natural. In the following,  we will propose two interesting choices of auxiliary random variables inspired by construction of auxiliary random variable for the network example. In the next section, we answer a more fundamental question: To what extent can  source correlation be captured by entropies.

\subsection{Linearly Correlated Random Variables}\label{sec:Linearly Correlated Random Variables}

In some scenarios, source random variables are ``linearly correlated''. In those cases, we can choose ``linear'' auxiliary random variables.
\begin{definition}
A set of random variables $\{Y_{1}, \ldots, Y_{n}\}$ is called linearly correlated if
\begin{enumerate}
\item for any $\alpha \subseteq \{1,\ldots, n\}$, the support of the probability distribution of $(Y_{i},i\in\alpha)$ is a vector subspace and
\item $(Y_{i},i\in\alpha)$ is uniformly distributed over its supports.
\end{enumerate}
\end{definition}

\begin{lemma}\label{lem:LinearlyCorrelated}
Let $(Y_{1}, \ldots, Y_{n})$ be a set of linearly correlated random variables with support vector subspaces 
$V_i \subseteq \mathbb{F}^{m}_{q}$ 
and 
\begin{equation}
\mathrm{dim} \langle V_i : i \in \{1,\ldots,n\} \rangle = m. \label{eq:SecLinear1}
\end{equation}
Let $(K_{1}, \ldots, K_{m})$ be the set of independent random variables uniformly distributed over the support $ {\mathbb F}_{q}$. Then there exists matrices $\mathbf{A}^i, i=1,\ldots,n$ such that
\begin{equation*}
Y_i = [K_1 \ldots K_m] \mathbf{A}^i
\end{equation*}
where 
\begin{equation*}
\mathbf{A}^i=
 \begin{bmatrix}
  a^i_{1,1} & a^i_{1,2} & \cdots & a^i_{1,\mathrm{dim}(V_i)} \\
  a^i_{2,1} & a^i_{2,2} & \cdots & a^i_{2,\mathrm{dim}(V_i)} \\
  \vdots  & \vdots  & \ddots & \vdots  \\
  a^i_{m,1} & a^i_{m,2} & \cdots & a^i_{m,\mathrm{dim}(V_i)}
 \end{bmatrix}
\end{equation*}
is an $m \times \mathrm{dim}(V_i)$ matrix.
\end{lemma}

\begin{IEEEproof}
Let $B(i)_1,\ldots,B(i)_{\mathrm{dim}(V_i)}$ be a basis for the vector subspace $V_i$ where $B(i)_1, \ldots,$  $B(i)_{\mathrm{dim}(V_i)}$ are column-$m$ vectors. Then we can construct a $m \times \mathrm{dim}(V_i)$ matrix $\mathbf{A}^i$ such that its column vectors are $B(i)_1,\ldots,B(i)_{\mathrm{dim}(V_i)}$. 

Note that, for any $(K_1,\ldots,K_m)=(k_1,\ldots,k_m)$,
\begin{equation*}
y_i=[k_1,\ldots,k_m]\mathbf{A}^i
\end{equation*}
is vector in the subspace $V_i$. Now we need to show that $y_i: y_i \in V_i$ are uniformly distributed. For any $Y=y_i$
\begin{equation*}
\mathrm{Pr}(Y_i=y_i)=\frac{|\{(k_1,\ldots,k_m):[k_1,\ldots,k_m]\mathbf{A}^i=y_i\}|}{q^m}.
\end{equation*}
But, for any $y_i$
$$|\{(k_1,\ldots,k_m):[k_1,\ldots,k_m]\mathbf{A}^i=y_i\}| =|\{(k_1,\ldots,k_m):[k_1,\ldots,k_m]\mathbf{A}^i=0\}|.$$
Hence the random variable $Y_i$ is equiprobable over its support vector subspace $V_i$. A similar argument can be made for any $(Y_{i}, i\in\alpha)$ for all subsets $\alpha $ of $\{1, \ldots, n\}$.
 \end{IEEEproof}
 
So, we proved that   $Y_1,\ldots,Y_n$ are linear functions of the random variables $K_{1}, \ldots,$ $K_{m}$. In particular,  $Y_i$ is a function of all the random variables $K_j$ where  the $j$th row of $\mathbf{A}^i$ is non-zero.

\subsection{Using Common Information}\label{sec:Using common information}
We can also construct an auxiliary random variable by using common information.
\begin{definition}[Common Information \cite{GacKor73}]
For any random variables $X$ and $Y$, the \emph{common information} of $X$ and $Y$ is the  random variable (denoted by $C(X,Y)$) which has the maximal entropy among all other random variables $K$ such that
\begin{align*}
    H(K|X) &=0\\
    H(K|Y) &=0.
\end{align*}
\end{definition}

There are cases where even though the random variables are highly correlated, their common information can still be small.  
For example, let $Z$ be a binary random variable such that $\Pr(Z=0)=\epsilon >0$ and $\Pr(Z=1) = 1-\epsilon$. 
Suppose $X$ is another binary random variable independent of $Z$ and $Y=X \oplus Z$. In this case, even if $X$ and $Y$ are highly correlated (when $\epsilon$ is small), their common information is still zero. In that case, we cannot choose the common information as the auxiliary random varible. To address this issue, we propose a different way to construct auxiliary random variables. 

Consider any pair of random variables $\{X,Y\}$ with probability distribution $ P_{XY}(\cdot)$. For any $\delta \ge 0$, let
\begin{equation*}
  \mathcal P(\delta) \triangleq \left\{ P_{K|XY}(\cdot):
  \begin{array}{l l}
     H(K|X) \leq \delta,\\
    H(K|Y) \leq \delta,\\
    I(X;Y|K) \leq \delta
  \end{array} \right\}
\end{equation*}
where the probability distribution of $\{X,Y,K\}$ is given by
$$\Pr(X=x,Y=y,K=k) \defined P_{XY}(x,y)P_{K|XY}(k|x,y).$$
Note that the ``smaller'' the $\delta$ is, the more similar the random variable $K$ (associated with the conditional distribution  $P_{K|XY}$) is to the common information.
Our constructed random variable can be selected from $\mathcal P(\delta^{*})$ to formulate an improved LP bound where
\begin{equation*}
\delta^{*} = \min_{\delta : \mathcal P(\delta) \neq \emptyset} \delta.
\end{equation*}

For a multi-source multicast network with source random variables $Y_1,\ldots,Y_{|\mathcal S|}$ one can construct random variables $K_{ij}, i \neq j, i,j \in \mathcal S$ from the family of distributions
\begin{equation*}
  \mathcal P(\delta) \triangleq \left\{ P_{K_{ij}|Y_i,Y_j}(\cdot):
  \begin{array}{l l}
     H(K_{ij}|Y_i) \leq \delta,\\
    H(K_{ij}|Y_j) \leq \delta,\\
    I(Y_i;Y_j|K_{ij}) \leq \delta
  \end{array} \right\}.
\end{equation*}

An improved LP bound for a multi-source multicast network with source random variables $Y_1,\ldots,Y_{|\mathcal S|}$ can be computed by using the auxiliary random variables $K_{ij}, i \neq j, i,j \in \mathcal S$.

\section{Distribution characterisation by entropies}\label{sec:DistributionViaEntropy}
In previous section, we have demonstrated how to use auxiliary random variables and entropic relations to capture the correlations among sources. A natural question then arises: can we completely capture the correlation using this method? More precisely, can one choose enough auxiliary random variables such that the joint probability distribution of the sources can be completely determined. 
In the following, we will show that the answer to the question is indeed affirmative.

To illustrate the idea, consider a random vector   $X=(X_{1}, \ldots, X_{M})$ with probability distribution $p_{X} (x_{1},\ldots, x_{M})$.  We can arbitrarily ``construct'' an auxiliary random variable $Y$ by specifying the conditional probability distribution $p_{Y|X} (y|x_{1},\ldots, x_{M})$. Now, instead of using only the entropy function defined in \eqref{1}, we can improve the ``quality'' of representation by using the ``extended entropy function'' 
\begin{align*}
h(W) \triangleq 
\begin{cases}
H(X_{s}  , s\in\alpha) & \text{ if } Y \not\in \alpha \\
H(Y, X_{s}  , s\in\alpha) & \text{ if } Y \in \alpha \\
\end{cases}
\end{align*}
for all subset of random variables  $W \subseteq \{X_{1}, \ldots X_{M}, Y\}$. 

For example, suppose one can construct an auxiliary random variable $Y$ such that 
\begin{align}\label{3}
H(Y|X_{1}) = H(Y|X_{2}) = 0 
\end{align} 
and 
\begin{align}\label{4}
H(Y) \ge \frac{1}{2} \max ( H(X_{1}), H(X_{2)}).
\end{align}
The conditions \eqref{3} and \eqref{4} already impose a very strong constraint on the joint probability distribution of $(X_{1},  X_{2})$ that 
$X_{1}$ and $X_{2}$ have a ``common information'' $Y$ of entropy at least half of the entropy of  each individual random variable.

The basic question now is: \emph{How ``accurate'' can entropy function specify the correlation among random variables}? We partly answer the question by 
showing that the joint probability distribution among random variables can be completely specified by entropy functions subject to cardinality constraint.
To understand why, consider a binary random variable $X$ such that $p_{X}(0) = p$ and 
$p_{X}(1) = 1-p$. While the entropy of $X$ does not determine exactly what the probabilities of $X$ are, it essentially determines the probability distribution (up to renaming).  To be precise, let $0\le q \le 1/2$  such that 
$
H(X) = h_{b}(q)
$
where
$
h_{b}(q) \triangleq -q\log q - (1-q) \log (1-q).
$
Then 
either $p=q$ or $p= 1-q$. Furthermore, the two possible distributions can be obtained from each other by renaming the random variable outcomes appropriately.

\def\PN{\{2,\ldots, m \}}

\def\X{{\cal X}}
\def\a{{\alpha}}

\subsection{Construction of Auxiliary Random Variables}\label{sec:Construction of Auxiliary Random Variables}

When $X$ is not binary, the entropy $H(X)$ alone is not sufficient to characterise the probability distribution of $X$. However, by using auxiliary random variables, it turns out that the distribution of $X$ can still be determined.  

The idea is best demonstrated by an example. Suppose $X$ is ternary, taking values from the set $\{1,2,3\}$. Suppose also that $p_{X}(x) > 0$ for all $x\in \{1,2,3\}$.
Define random variables $A_{1}$, $A_{2}$ and $A_{3}$ such that 
\begin{align}\label{2b}
A_{i} = 
\begin{cases}
1 & \text{ if } X = i \\
0 & \text{ otherwise. }
\end{cases}
\end{align}
Clearly, 
\begin{align}\label{2a}
H(A_{i}|X) = 0
\end{align}
 and 
\begin{align}\label{2}
H(A_{i}) = h_{b}(p_{X}(i)).
\end{align}
Let us further assume that $p_{X}(i) \le 1/2$ for all $i$. Then by \eqref{2} and strict monotonicity of 
$h_{b}(q)$ in the interval $[0, 1/2]$, it seems at the first glance that the distribution of $X$ is uniquely specified by the entropies of the auxiliary random variables. 

However,  there is a catch in the argument -- The auxiliary random variables chosen are not arbitrary. When we ``compute'' the probabilities of $X$ from the entropies of the auxiliary random variables, it is assumed that how the random variables are constructed is known. Without knowing the ``construction'',  it is unclear how to find the distribution of $X$ from entropies.

More precisely, suppose we only know that there exists auxiliary random variables $A_{1},A_{2},A_{3}$ such that \eqref{2a} and \eqref{2} hold (without knowing that the random variables are constructed according to \eqref{2b}). Then in this case, we cannot possibly determine precisely what the distribution of $X$ is.  
Despite the difficulties, we will show  how to construct auxiliary random variables via which the probability distribution can be characterised from entropies.

Let $X$ be a random variable  with support ${\cal N}_{n}=\{1,\ldots,n\}$ and $\Omega$ be the set of all nonempty binary partitions of ${\cal N}_{n}$. In other words, $\Omega$ is 
the collection of all sets $\{\alpha, \alpha^{c}\}$ such that $\alpha \subseteq {\cal N}_{n}$,  and both $|\alpha| $ and $|\alpha^{c}|$ are nonzero.
We will use $\lr \alpha \rr$ to denote the set $\{\alpha, \alpha^{c}\}$. 
To simplify notations, we may assume without loss of generality that $  \alpha $ is  a subset of $  \{2,\ldots,n\} $.
 Clearly, $|\Omega| = 2^{n-1}-1$.
Unless explicitly stated otherwise,  we may assume without loss of generality that 
 the probability that $X=i$ (denoted by $p_i$) is monotonic decreasing. In other words, 
$$p_1 \geq \ldots \geq p_n>0.$$

\begin{definition}[Partition Random Variables]\label{def:RVsAalpha}
For any random variable $X$ with support  ${\cal N}_{n}$, it induces $2^{n-1}-1$ binary random variables
$\{A_{\lr \alpha \rr} : \: \alpha \in \Omega \}$ such that 
\begin{equation*}
  A_{\lr \alpha \rr} \triangleq \left\{
  \begin{array}{l l}
    \alpha & \quad \text{if $X \in \alpha$}\\
    \alpha^{c} & \quad \text{otherwise.}\\
  \end{array} \right.
\end{equation*}

We call $\{A_{\lr \alpha \rr}, \alpha \in \Omega \}$ the collection of \emph{binary partition random variables} of $X$.
\end{definition}

\begin{remark}
If $|\alpha| = 1 $ or $n-1$,   then there exists an element $i\in \X$ such that $A_{\lr \alpha \rr} = \{i\}$ if and only if $X= i$. Hence, $A_{\lr \alpha \rr}$ is essentially a binary variable \emph{indicating/detecting whether $X = i$ or not}. In that case, we call $A_{\lr \alpha \rr}$ an \emph{indicator random variable}. 
Furthermore, when $n\ge 3$, there are exactly $n$ indicator variable, one for each element in ${\cal N}_{n}$. As we shall see,  if we can have the entropies of all the  indicator random variables, then we can determine all  the probabilities $p_i, i = 1,\ldots,n$. 
\end{remark}

In the following we show that, given a set of auxiliary random variables without knowing how they are constructed, it is possible to determine if they are binary partition random variables or even indicator random variables. To achieve this goal, we first need to know some basic properties of the set of all binary partition random variables.

\begin{lemma}[Properties]\label{lemma:secaux1}
Let $X $ be a random variable with support  ${\cal N}_{n}$,  and  $(A_{\lr \alpha \rr}, \: \alpha \in \Omega)$ be its induced binary partition random variables. Then the following properties hold:
\begin{enumerate}
\item (Distinctness) for any $\lr\alpha\rr \neq \lr\beta\rr$, 
\begin{align}
H(A_{\lr \alpha \rr}|A_{\lr \beta \rr})&>0 \label{eq:lemma1a}\\
H(A_{\lr \beta \rr}|A_{\lr \alpha \rr})&>0.\label{eq:lemma1b}
\end{align}

\item (Completeness)
Let $A^{*}$ be a binary random variable such that 
$H(A^{*} | X) = 0$ and $H(A^{*} ) > 0$. Then there exists $\lr\alpha \rr \in \Omega$ such that
\begin{align*}
H(A^{*} | A_{\lr \alpha \rr}) = H( A_{\lr \alpha \rr} | A^{*} ) = 0.
\end{align*}
In other words, $A_{\lr \alpha \rr}$ and $A^{*}$ are essentially the same random variable.

\item (Basis) Let ${\lr \alpha \rr} \in \Omega$. Then there exists 
\[
\lr \beta_{1} \rr , \ldots , \lr \beta_{n-2}  \rr \in \Omega 
\]
such that 
\begin{align}
H(A_{\lr\beta_k\rr}|A_{\lr \alpha \rr},A_{\lr\beta_1\rr},\ldots,A_{\lr\beta_{k-1} \rr})&>0 \label{eq:secAux5}
\end{align}
for all $k=1, \ldots, n-2$. 
\end{enumerate}
\end{lemma}
\begin{IEEEproof}
See Appendix \ref{app.A}.
\end{IEEEproof}


\begin{proposition}[Characterising indicators]\label{prop1}
Let $X $ be a random variable of support ${\cal N}_{n}$ where $ n \ge 3$. Consider the binary partition random variables induced by $X$. Then for  all $i \ge 2$,
\begin{enumerate}
\item   
$
H(A_{\lr i\rr}|A_{\lr j \rr}, j > i) >0 
$, and 

\item For all 
$\alpha \in \Omega$ such that   $H(A_{\lr \alpha \rr}|A_{\lr j \rr} , j>i )> 0$, we have 
\begin{align}
H(A_{\lr i \rr}) &\leq H(A_{\lr \alpha \rr}). \label{eq:secAux9}
\end{align}

\item
Equalities \eqref{eq:secAux9} hold if and only if 
$A_{\lr \alpha \rr}$ is an indicator random variable detecting an element $\ell \in {\cal N}_{n}$  
such that 
$$p_{\ell} = p_{i}.$$  

\item 
If $A_{\lr \alpha \rr}$ is a binary partition random variable such that  
\[
H(A_{\lr \alpha \rr} | A_{\lr j \rr}, j \in \beta ) > 0
\]
for all proper subset $\beta$ of $[2,n]$\footnote{For integers $a,b$, we use $[a,b]$ to denote the set $\{a, a+1, \ldots, b\}$ in this paper.}, then $A_{\lr \alpha \rr} = A_{\lr 1 \rr}$.

\end{enumerate}
\end{proposition}
\begin{IEEEproof}
See Appendix \ref{app.A}.
\end{IEEEproof}

\def\N{{\cal N}}  

In Proposition \ref{prop1}, we have obtained various properties about the indicator random variables. In the following, we will show that by using the binary partition random variables (and their entropies), one can characterise the probability distribution of a random variable. As we shall see, the proof of this result is based on the properties of indicator random variables. 
  
\begin{theorem}[Random Scalar Case]\label{thm1}
Suppose $X$ is a random variable with support $\N_{n}$.   
For any $\lr\alpha\rr \in \Omega$, let $A_{\lr \alpha \rr}$ be the corresponding binary partition random variables. 
Now, suppose  $X^{*}$ is another  random variable  such that 
\begin{enumerate}
\item
 the size of its support $\X^{*}$ is at most the same as that of $X$, and 
 
 \item  there exists random variables $(B_{\lr \alpha \rr}, \lr \alpha \rr \in\Omega)$ satisfying the following conditions:
\begin{align}
H(B_{\lr \alpha \rr}, \alpha \in \Delta) & = H(A_{\lr \alpha \rr},\lr \alpha \rr\in\Delta), \label{eq34}\\
H( B_{\lr \alpha \rr} | X^{*}) & =0 \label{eq35}
\end{align}
for all $\Delta \subseteq \Omega$.  
\end{enumerate}

Then, the following 
properties hold: 
\begin{enumerate}
\item (\emph{Distinctness}) All the random variables $B_{\lr \alpha \rr}$ for $\lr \alpha \rr\in\Omega$ are distinct and have non-zero entropies.

\item (Basis) Let $\lr\alpha\rr  \in \Omega$. Then there exists 
\[
\lr\beta_{1}\rr , \ldots , \lr\beta_{n-2}\rr \in \Omega 
\]
such that 
\begin{align*}
H(B_{\lr\beta_k\rr}|B_{\lr \alpha \rr},B_{\lr\beta_1\rr},\ldots,B_{\lr\beta_{k-1}\rr})&>0  
\end{align*}
for all $k=1, \ldots, n-2$. 

\item  (\emph{Binary properties})
For any $\lr\alpha\rr\in\Omega$, $B_{\lr \alpha \rr}$ is a binary partition random variable of $X^{*}$.   In this case,  we may assume without loss of generality that there exists   $\omega_{\lr \alpha \rr} \subseteq \X^{*}$ such that  
\begin{equation*} 
  B_{\lr \alpha \rr} = \left\{
  \begin{array}{l l}
    \omega_{\lr \alpha \rr} & \quad \text{if $X^{*} \in \omega_{\lr \alpha \rr}$}\\
    \omega_{\lr \alpha \rr}^{c} & \quad \text{otherwise}\\
  \end{array} \right.
\end{equation*}

\item (Completeness)
Let $B^{*}$ be a binary partition random variable of $X^{*}$ with non-zero entropy. Then there exists $\lr\alpha\rr \in \Omega$ such that
\begin{align*}
H(B^{*} | B_{\lr \alpha \rr}) = H(B_{\lr \alpha \rr} | B^{*} ) = 0.
\end{align*}

\item (\emph{Indicator})
If $|\alpha| = \{i\}  $  for any $i=1,\ldots, n$,  then 
$B_{\lr \alpha \rr}$ is an \emph{indicator variable}.

\item (\emph{Distribution equivalence})
There is a mapping 
$$
\sigma : \N_{n} \to \X^{*}
$$
such that 
$
\Pr(X=i) = \Pr(X^{*} = \sigma(i)).
$
In other words, the probability distributions of $X$ and $X^{*}$ are essentially the same (via renaming outcomes).
\end{enumerate}
\end{theorem}
\begin{IEEEproof}
See Appendix \ref{app.B}.
\end{IEEEproof}

\begin{remark}
Note that, given a set of random variables satisfying \eqref{eq34} and \eqref{eq35}, i.e., partition random variables, for a given scalar random variable, it is feasible to obtain probability distribution (up to relabeling) of the random variable via entropy functions of the indicator random variables. 
\end{remark}

In the following, we will extend Theorem \ref{thm1} to the case of random vector. Such extension  is not as trivial as it may seem, as illustrated in the following example.

\begin{example}
Consider two random vectors  $X =(X_{1}, X_{2})$ and $X^{*} =(X^{*}_{1}, X^{*}_{2})$ with probability distributions given in Table \ref{table1}.
\begin{table}[]
\centering
\caption{Probability distributions of $X$ and $X^{*}$}
\label{table1}
\begin{tabular}{cccccc}
  &                        &                          & $X_{2}$                        &                          &                          \\
  &                        & $1$                        & $2$                        & $3$                        & $4$                        \\ \cline{3-6} 
  & \multicolumn{1}{c|}{$a$} & \multicolumn{1}{c|}{$1/8$} & \multicolumn{1}{c|}{$1/8$} & \multicolumn{1}{c|}{$0$}    & \multicolumn{1}{c|}{$0$}    \\ \cline{3-6} 
$X_{1}$ & \multicolumn{1}{c|}{$b$} & \multicolumn{1}{c|}{$1/8$} & \multicolumn{1}{c|}{$1/8$} & \multicolumn{1}{c|}{$0$}    & \multicolumn{1}{c|}{$0$}    \\ \cline{3-6} 
  & \multicolumn{1}{c|}{$c$} & \multicolumn{1}{c|}{$0$}    & \multicolumn{1}{c|}{$0$}    & \multicolumn{1}{c|}{$1/8$} & \multicolumn{1}{c|}{$1/8$} \\ \cline{3-6} 
  & \multicolumn{1}{c|}{$d$} & \multicolumn{1}{c|}{$0$}    & \multicolumn{1}{c|}{$0$}    & \multicolumn{1}{c|}{$1/8$} & \multicolumn{1}{c|}{$1/8$} \\ \cline{3-6} 
%
&&&&\\
%
  &                        &                          & $X^{*}_{2}$                        &                          &                          \\
  &                        & $1$                        & $2$                        & $3$                        & $4$                        \\ \cline{3-6} 
  & \multicolumn{1}{c|}{$a$} & \multicolumn{1}{c|}{$1/8$} & \multicolumn{1}{c|}{$1/8$} & \multicolumn{1}{c|}{$0$}    & \multicolumn{1}{c|}{$0$}    \\ \cline{3-6}
$X^{*}_{1}$ & \multicolumn{1}{c|}{$b$} & \multicolumn{1}{c|}{$0$} & \multicolumn{1}{c|}{$1/8$} & \multicolumn{1}{c|}{$1/8$}    & \multicolumn{1}{c|}{0}    \\ \cline{3-6} 
  & \multicolumn{1}{c|}{$c$} & \multicolumn{1}{c|}{$0$}    & \multicolumn{1}{c|}{$0$}    & \multicolumn{1}{c|}{$1/8$} & \multicolumn{1}{c|}{$1/8$} \\ \cline{3-6} 
  & \multicolumn{1}{c|}{$d$} & \multicolumn{1}{c|}{$1/8$}    & \multicolumn{1}{c|}{$0$}    & \multicolumn{1}{c|}{$0$} & \multicolumn{1}{c|}{$1/8$} \\ \cline{3-6} 
\end{tabular}
\end{table}
If we compare the joint probability distributions of 
$X $ and $X^{*}$, they are different from each other. Yet, if we treat  $X $ and $ X^{*} $ as scalars (by properly renaming), then they indeed have the same distribution (both uniformly distributed over a support of size 8). This example shows that we cannot directly applying Theorem \ref{thm1} to the random vector case, by simply mapping a vector into a scalar.
\end{example}

\begin{theorem}[Random Vector Case]\label{thm2}
Let  $X$ be  a random vector $(X_{1},\ldots, X_{M})$ with support $\X $. 
Let $\Omega$ be the set of all nonempty binary partitions of $\X$ and  $A_{\lr \alpha \rr}$ be the binary partition random variable of $X$ such that 
\begin{equation*}
  A_{\lr \alpha \rr} = \left\{
  \begin{array}{l l}
    \alpha & \quad \text{if $X \in \a$}\\
    \alpha^{c} & \quad \text{otherwise} 
  \end{array} \right.
\end{equation*}
 for all $\lr \alpha \rr \in \Omega$.

Now, suppose  $X^{*} = (X^{*}_{1} , \ldots, X^{*}_{M})$ is another random vector  where  there exists random variables 
$$
(B_{\lr \alpha \rr}, \lr\alpha\rr \in\Omega)
$$ 
such that for any subset $\Delta$ of $\Omega$ and $\tau \subseteq \{1, \ldots, M\}$,
\begin{align}\label{thm5:eqa}
H(B_{\lr \alpha \rr}, \lr\alpha\rr \in \Delta, X^{*}_{j}, j \in \tau) = H(A_{\lr \alpha \rr}, \lr\alpha\rr \in \Delta, X_{j}, j \in \tau). 
\end{align}
 %
 Then the joint probability distributions of $X=(X_{1},\ldots, X_{n})$ and $X^{*} = (X^{*}_{1}, \ldots, X^{*}_{n})$ are the same (subject to relabelling). More precisely,  there exists bijective mappings $\sigma_{m}$  for $m=1,\ldots, M$ such that 
$$\Pr (X = (x_{1}, \ldots , x_{M}) ) = \Pr (X^{*} = (\sigma_{1}(x_{1}), \ldots , \sigma_{M}(x_{M}) )). $$
\end{theorem}
\begin{IEEEproof}
See Appendix \ref{app.C}.
\end{IEEEproof}

\begin{remark}
Note that, from Theorem \ref{thm2}, it is feasible to obtain probability distribution (up to relabeling) of the vector random variable via the entropy functions of a set of random variables satisfying \eqref{thm5:eqa}. 
\end{remark}

\section{Extensions}\label{sec:Extensions}
In this paper, we proposed the use of auxiliary random variables (and their entropies) to characterise the correlations among sources. As a result, we can sharpen the LP bound for network coding. In the following, we will outline a few ideas of how to further extend our work. 

First, we can extend how to define auxiliary random variables. The framework we proposed earlier can be viewed as ``symbol characterisation''. Roughly speaking, we treat the sources as i.i.d. copies of a vector of correlated source symbols. Yet, we can naturally extend the framework to ``block characterisation'' by considering i.i.d. copies of a vector of  source blocks (of symbols). In other words, for each source $s$, we consider a super source symbol corresponding to a block of source symbols
\[
(Y_{s}^{1} , \ldots, Y_{s}^{m}).
\]
Here, $m$ is the block length. When $m=1$, it reduces to the scenario we described in the beginning.

Under this extension, our work and the work in \cite{GohYanJagg13} (see also \cite{GohYanJagg11}) shared some similarities. In \cite{GohYanJagg13}, the authors 
proposed a new method to improve the cut-set bound for networks with correlated sources. Their idea was based on the use of ``common information'' and cut-set bound. Suppose there are two sources $Y_{s}^{(N)} \defined (Y_s^n, n = 1 \ldots N)$ for $s \in \{1,2\}$. The authors aimed to construct an auxiliary random variable $K$ 
where  the random variable $K$ may be interpreted as ``almost'' common information of the two sources. Specifically, 
$$
\frac{H(K|Y_1^{(N)})}{N},\frac{H(K|Y_2^{(N)})}{N},\frac{H(K|Y_1^{(N)},Y_2^{(N)})}{N}
$$
are chosen to be  as small as possible, while 
$$
\frac{H(K)}{N}
$$
as large as possible.
By using $K$, the authors can tighten the cut-set bound. The characterization of  ``uncertainty region'' is single-letter and the authors also described its application to bounding the network capacity of secure transmission in the presence of an eavesdropper. Alternatively, one can view that \cite{GohYanJagg13} proposed to loosely decouple the (block of) sources into three parts $X_{1}, X_{2}, K$ such that 
 \begin{align*}
 Y_1^{(N)} &= (X_{1}, K) \\
 Y_2^{(N)} &= (X_{2}, K)  
 \end{align*}
and the three parts are treated as mutually independent. In this sense, the spirit of  \cite{GohYanJagg13} and the construction of auxiliary random variables in Section \ref{sec:Using common information}  are similar. However, in our framework, we are not limited to auxiliary random variables corresponding to common information. We investigated a more general question: is it feasible to characterise probability distribution (or source correlation) completely using entropy functions? As a result, our characterisation of distributions via entropies in Section \ref{sec:DistributionViaEntropy} can provide stronger converse results since the ``uncertainty region'' can be obtained by the joint distribution of source random variables but the converse may not be true in general. It should also be noted that the approach of designing auxiliary random variables described in \cite{GohYanJagg13} as well as in this paper are not only applicable to improve cut-set type bounds but are also equally useful to improve geometric bounds. Here, we are using geometric bounds, instead of the cut-set bound (which is a relaxation of the LP bound).

Second, Theorem \ref{thm2} showed that  one can use entropies of auxiliary random to completely characterise the joint probability distribution of a random vector. In this paper, Shannon entropies are implicitly referred to. However, it can be verified easily that the same results hold for other entropies including R\'{e}nyi entropies (of order $\alpha >0$) \cite{Ren60} and Tsallis entropies \cite{Tsa88}. More specifically, let $H(X)$ be an entropy measure satisfying the following two properties: 
\begin{enumerate}
\item (Monotonicity)
Let  $X$ be a binary random such that $p_{0} = p \le 1/2$, and $h(p)$ be  its  entropy. Then 
$h$ is a strictly increasing function of $p$ between $[0,1/2]$.

\item (Functional dependency)
\[
H(X)  = H(X,Y)
\]
if and only if $Y$ is a function of $X$.

\end{enumerate}

As long as these two properties are satisfied, then the entropies of the auxiliary random variable (constructed in Theorem \ref{thm2}) will be sufficient to uniquely characterise the probability distribution of a set of random variables.


\section{Conclusion}\label{sec:conclusion}

In this paper, we have considered outer bounds for network coding capacity when sources are correlated. We proposed the use of auxiliary random variables to  better capture the source correlations, leading to tighter outer bounds for the achievable region. We also showed that by using auxiliary random variables, entropic relations are sufficient to uniquely characterise the  probability distribution of a random vector (up to relabeling). 
Yet, there are many open questions remained to be answered. For example, the proposed construction of the auxiliary random variables is not optimised in any sense. 
Suppose we can only use only a fixed number of auxiliary random variables, how well  entropies  can represent the correlation among random variables?  This question is still unanswered.

\begin{appendices}

\section{Partition induced random variables}\label{app.A}

In this appendix, we will prove some interesting properties of partition random variables.

\begin{lemma}\label{lemma:secaux3}
Let $D$ be a random variable over a support of size at most  $n$ and   $C_1,\ldots,C_{n-1}$ be functions of $D$. In other words,
$H(C_{i}| D) = 0$ for all $i=1, \ldots, n-1$.
Then the following two statements are equivalent:
\begin{enumerate}
\item for all $i=1,\ldots,n-1$, 
\begin{align}
H(C_i|C_{i-1},\ldots,C_1)>0.  \label{eq:secAux1}
\end{align}

\item
$\mathcal S(C_1,\ldots,C_i)=i+1$, where $\mathcal S(C_1,\ldots,C_i)$ is the size of the support of $(C_1,\ldots,C_i)$.

\end{enumerate}
\end{lemma}
\begin{IEEEproof}[Proof of Lemma \ref{lemma:secaux3}] 
The key to the proof rests on the following simple observation: 
For any random variables $X$ and $Y$,  $H(Y|X)> 0 $ if and only if $ \mathcal S(X,Y)>\mathcal S(X)$. 
Following the observation, we can easily prove that 
inequalities \eqref{eq:secAux1} hold  for all $i=1,\ldots,n-1$ if and only if 
\begin{align}
2 \le \mathcal S(C_1) < \ldots <\mathcal S(C_1,\ldots,C_{n-1}). \label{eq:secAux2}
\end{align}
Together with the assumption that  
 $\mathcal S(C_1,\ldots,C_{n-1}) \le \mathcal S(D) \le n$, 
the lemma is proved.
\end{IEEEproof}

\subsection{Proof of Lemma \ref{lemma:secaux1}}
First, we prove the distinctness property. Note that 
\begin{align*}
(A_{\lr \alpha \rr},A_{\lr \beta \rr}) = 
\begin{cases}
(\alpha^{c},\beta^{c}) & \text{ if } X \not\in \alpha \cup \beta \\
(\alpha,\beta^{c}) & \text{ if } X  \in \alpha \setminus \beta \\
(\alpha^{c},\beta) & \text{ if } X  \in \beta \setminus \alpha  \\
(\alpha,\beta) & \text{ if } X  \in \alpha \cap \beta.
\end{cases}
\end{align*}
Follow our convention, we assume that  $1 \not\in \alpha \cup \beta$. Hence, 
\begin{align*}
\Pr(A_{\lr \alpha \rr}=\alpha^{c}, A_{\lr \beta \rr}=\beta^{c}) & \ge \Pr(X=1)   > 0.
\end{align*}

Next, since $\alpha $  is nonempty, either $\alpha \cap \beta  $ or 
$\alpha \setminus \beta $  are nonempty. Similarly, as $\beta$ is nonempty, either 
$\alpha \cap \beta  $ or $\beta \setminus \alpha $  are nonempty. 
Suppose $\alpha \cap \beta = \emptyset$. Then both $\alpha \setminus \beta $ and $\beta \setminus \alpha $  are nonempty. Consequently, 
\begin{align*}
\Pr(A_{\lr \alpha \rr}=\alpha^{c}, A_{\lr \beta \rr}=\beta) & = \Pr(X \in \beta\setminus \alpha)  > 0
\end{align*}
and 
\begin{align*}
\Pr(A_{\lr \alpha \rr}=\alpha, A_{\lr \beta \rr}=\beta^{c}) & = \Pr(X \in \alpha\setminus \beta)  > 0 .
\end{align*}
In this case, it is obvious that \eqref{eq:lemma1a} and \eqref{eq:lemma1b} hold.

Now, suppose $\alpha \cap \beta \neq \emptyset$ and hence 
\begin{align*}
\Pr(A_{\lr \alpha \rr}=\alpha, A_{\lr \beta \rr}=\beta) & = \Pr(X \in \alpha\cap\beta)  > 0 .
\end{align*}
Since  $\alpha \neq \beta $, either $\alpha\setminus \beta \neq \emptyset$ or $\beta \setminus \alpha \neq \emptyset$. In other words, either 
 \begin{equation*}
\Pr(A_{\lr \alpha \rr}=\alpha, A_{\lr \beta \rr}=\beta^{c}) = \Pr(X\in \alpha\setminus \beta) >0
\end{equation*}
or
\begin{equation*}
\Pr(A_{\lr \alpha \rr}=\alpha^{c}, A_{\lr \beta \rr}=\beta) = \Pr(X\in \beta\setminus \alpha) >0.
\end{equation*}
Again, this implies that  
\begin{align*}
H(A_{\lr\alpha\rr}|A_{\lr \beta \rr})&>0\\
H(A_{\lr \beta \rr}|A_{\lr \alpha \rr})&>0.
\end{align*}
We proved the distinctness property. 

To prove the completeness property, let $A^{*}$ be a binary random variable such that 
$H(A^{*} | X) = 0$ and $H(A^{*} ) > 0$. 
Since  $A^{*} $ is binary, there exists a nonempty proper subset $\beta$ of $  \{1, \ldots, n\}$ such that 
\begin{equation*}
A^{*}= \left\{
\begin{array}{l l}
   1 & \quad \text{if $X \in \beta$}\\
    0 & \quad \text{otherwise}\\
  \end{array} \right.
\end{equation*}
Clearly, $A^{*}$ and $A_{\lr\beta\rr}$ are the same in the sense that 
$$
H(A^{*} | A_{\lr\beta\rr}) = H(A_{\lr\beta\rr}|A^{*}) = 0.
$$
%

Finally, we will prove the basis property. 
Let  $\lr\alpha\rr  \in \Omega$. Assume without loss of generality that $\alpha=\{i,\ldots,n\}$. Let
\begin{align*}
\beta_{1} = \{2\}, \ldots, \beta_{n-2}  = \{n-1\}.
\end{align*}
We can directly verify that 
$$\mathcal S(A_{\lr \alpha \rr})<\mathcal S(A_{\lr \alpha \rr}, A_{\lr\beta_{1}\rr})<\ldots <\mathcal S(A_{\lr \alpha \rr}, A_{\lr\beta_{1}\rr},\ldots,A_{\lr\beta_{n-2}\rr}).$$
Invoking Lemma \ref{lemma:secaux3}, we have 
\begin{align*}
H(A_{\lr\beta_k\rr}|A_{\lr \alpha \rr},A_{\lr\beta_1\rr},\ldots,A_{\lr\beta_{k-1}\rr}) &>0
\end{align*}
for all $k=1,\ldots,n-1$. The lemma thus follows.

\subsection{Proof of Proposition \ref{prop1}}

Let $\lr\alpha\rr \in \Omega$. Following the convention that $1\not\in\alpha$,  it can be proved directly that 
 $H(A_{\lr \alpha \rr}|A_{\lr j\rr}, j > i)> 0$ if and only if 
 $
\alpha \setminus [i+1,n] \neq \emptyset
$.   
Hence, we proved 1).

Next, notice that the binary entropy function 
$$
h_{b}(x) \triangleq -x \log x - (1-x) \log (1-x)
$$
is concave and symmetric at $x=0.5$. Hence, 
\begin{align*}
h_{b}(x) & = \hb{0.5 -  \left| x - 0.5 \right|    }.
\end{align*}
Next, it can be verified directly that 
\begin{align*}
 \hspace{-1cm} H(A_{\lr \alpha \rr} )  
& =   h_{b}\left(\sum_{k\in\alpha   }p_{k}  \right) \nonumber\\
& =    h_{b}\left( 0.5 - \left| \sum_{k\in\alpha  }p_{k}  - 0.5 \right| \, \right). 
\end{align*}
Hence, 
\begin{align*}
H(A_{\lr i \rr} ) =   h_{b}(p_{i} ) > 0.
\end{align*}

By definition, 
$p_{i} \le p_{k}$ for all $k \le i $. Hence, 
\begin{align*}
p_{i}  - 0.5 & \le \sum_{k\in\alpha \setminus [i+1,n] } p_{k}  - 0.5  \\
& \le \sum_{k\in\alpha } p_{k}  - 0.5.
\end{align*}
On the other hand, as $p_{i} \le p_{1}$. Hence,
\begin{align*}
 {\sum_{k\in\alpha  }p_{k} + p_{i } }  & \le  {\sum_{k\in\alpha}p_{k} + p_{1 }}  \\
& \le 1.
\end{align*}
Consequently,
\begin{align*}
p_{i} - 0.5\le  \sum_{k\in\alpha  }p_{k} - 0.5  \le 0.5 - p_{i} 
\end{align*}
or equivalently, 
\begin{align*}
\left|
 \sum_{k\in\alpha}p_{k}   - 0.5
  \right| \le  |0.5 - p_{i} |.
\end{align*}

As $\hb{x}$ is a strictly increasing function for $0 \le x \le 0.5$,  we have 
\begin{align*}
H(A_{\lr i \rr}) \le H(A_{\lr\alpha\rr})
\end{align*}
and thus prove 2).

Also, equality holds if and only if 
\[
\left|
 \sum_{k\in\alpha  }p_{k}   - 0.5
  \right| =  |0.5 - p_{i} |
\]
which is equivalent to either
\begin{align}\label{case1}
\sum_{k\in\alpha} p_{k}   =    p_{i}   
\end{align}
or
\begin{align}\label{case2}
\sum_{k\in\alpha} p_{k}   =  1 - p_{i} .
\end{align}
When \eqref{case1} holds, this means that 
 $\alpha = \{ \ell \}$ for some $\ell \in  [2,i]$  and $p_{\ell} = p_{i}$. 
On the other hand, when \eqref{case2} holds, this means that 
\[
p_{i} + \sum_{k\in\alpha} p_{k} =1. 
\]
Hence,  $\alpha = [2,n]$ and $p_{1} = p_{i}$.
In any case, $A_{\lr\alpha\rr}$ is an indicator variable for an element $\ell$ such that $p_{\ell} = p_{i}$. We thus prove 3).

Finally, 4) can be proved by direct verification.  
The proposition is thus proved. 

\section{Proof of Theorem \ref{thm1} - Random Scalar Case}\label{app.B}
Consider a random variable $X^{*}$ whose support is $\X^{*}$ of size at most $n$. 
If $n=2$, we already know that the distribution is uniquely determined by the random variable's entropy. Therefore, we will assume that  $n\ge 3$ in the following. For simplicity, we may assume that 
$\X^{*}$ is a subset of $\N_{n}$ and $X^{*}$  has probability masses
\[
q_{1} \ge q_{2} \ge \ldots \ge q_{n} \ge 0. 
\]
In this case, we will not assume that $q_{n}>0$. However, as we shall see, $q_{n}$ is indeed positive. 
Now, let 
$
(B_{\lr \alpha \rr} , \alpha \in \Omega)
$
be a set of random variables as defined in Theorem \ref{thm1}, restated as below:
\begin{align*}
H(B_{\lr \alpha \rr}, \lr \alpha \rr \in \Delta)&=H(A_{\lr \alpha \rr}, \lr \alpha \rr \in \Delta), \quad \forall \Delta \subseteq \Omega \\  
H(B_{\lr \alpha \rr}|X^{*}) &=0,  \quad \forall \lr \alpha \rr \in \Omega.
\end{align*}

Consequently, from Lemma \ref{lemma:secaux1}, for all distinct $\lr\alpha\rr, \lr\beta\rr \in \Omega$, 
\begin{align*}
H(B_{\lr \alpha \rr} | B_{\lr \beta \rr}) = H(A_{\lr \alpha \rr} | A_{\lr \beta \rr}) &> 0 \\
H(B_{\lr \beta \rr} | B_{\lr \alpha \rr}) =H(A_{\lr \beta \rr} | A_{\lr \alpha \rr}) &> 0. 
\end{align*}
The distinctness property then follows. Similarly,  the basis properties follow from the basis properties in Lemma \ref{lemma:secaux1}.

Invoking Lemma \ref{lemma:secaux3} and the basis properties, we have
$$2 \le \mathcal S(B_{\lr \alpha \rr})<\mathcal S(B_{\lr \alpha \rr},B_{\lr\beta_1\rr})<\ldots<\mathcal S(B_{\lr \alpha \rr},B_{\lr\beta_1\rr},\ldots,B_{\lr\beta_{n-1}\rr})$$
%
Since  
$$
\mathcal S(B_{\lr \alpha \rr},B_{\lr\beta_1\rr},\ldots,B_{\lr\beta_{n-1}\rr})\leq \mathcal S(X^{*}) \le n,$$
 $\mathcal S(B_{\lr \alpha \rr}) = 2$. In other words,  $B_{\lr \alpha \rr}$ is a  binary random variables and the binary property is proved. 

Next, we will prove the completeness property. As $\mathcal S(B) \le n$, there are at most $2^{n-1}-1$  distinct binary random variables. By the distinctness property, 
all the variables $(B_{\lr \alpha \rr} , \lr\alpha\rr \in \Omega)$ are distinct. The result then follows.
In fact, we proved that the support size of $X^{*}$  is $n$ and hence $q_{n} >0$.


So far,  we have proved that $B_{\lr \alpha \rr}$ is a binary random variable. 
Therefore, we may assume without loss of generality that there exists  
$\omega_{\lr \alpha \rr} \subseteq \X^{*}$ such that  
\begin{equation*} 
  B_{\lr \alpha \rr} = \left\{
  \begin{array}{l l}
   \omega_{\lr \alpha \rr}  & \quad \text{if $X^{*} \in \omega_{\lr \alpha \rr}$}\\
   \omega_{\lr \alpha \rr}^{c}  & \quad \text{otherwise.} 
  \end{array} \right.
\end{equation*}
Let $A^{*}_{\lr \alpha \rr}$ be the set of partition random variables induced by $X^{*}$.
Hence, we have 
\[
B_{\lr \alpha \rr} = A^{*}_{\lr\omega_{\lr \alpha \rr}\rr}
\]

Now, we will prove the \emph{indicator property} recursively. 
Consider the base  case when $i=n$.
From Proposition \ref{prop1} and \eqref{eq34}-\eqref{eq35},
\begin{align*} 
H(B_{\lr n\rr})  = H(A_{\lr n\rr}) \le  H(A_{\lr \alpha \rr}) = H(B_{\lr \alpha \rr})
\end{align*}
for all $\lr\alpha\rr \in \Omega$.
Therefore, 
\begin{align*} 
H(A^{*}_{\lr \omega_{\lr n\rr} \rr}) \le  H(A^{*}_{\lr \omega_{\lr \alpha \rr}\rr})
\end{align*}
for all $\lr\alpha\rr \in \Omega$.
Together with  Proposition \ref{prop1}, this further implies that 
\begin{align*} 
H(A^{*}_{\lr \omega_{\lr n\rr} \rr}) = H(A^{*}_{\lr n\rr}).
\end{align*}

Invoking Proposition \ref{prop1} again, we prove that 
$A^{*}_{\lr \omega_{\lr n \rr} \rr }$ (and hence $B_{\lr n \rr}$) is an indicator random variable for an element $\ell  \in {\cal N}_{n}$ such that $q_{\ell} = q_{n}$. 
By renaming the elements properly, we may  assume without loss of generality that $B_{\lr n \rr} = A^{*}_{\lr n\rr}$.
 
Now, assume that 
$
B_{\lr k \rr} = A^{*}_{\lr k\rr}
$ (subject to relabelling) 
for $k \ge i$ and $i \ge 3$. We will now prove that, subject to element renaming,  
\[
B_{\lr i-1 \rr} = A^{*}_{\lr i-1 \rr}.
\] 

First, by \eqref{eq34}-\eqref{eq35}, we have
$
H(B_{\lr \alpha \rr}| B_{\lr j \rr}, j > i -1)   > 0
$ 
 if and only if 
$
H(A_{\lr \alpha \rr}| A_{\lr j \rr}, j > i-1)   > 0
$.  
Hence,  
$$
H(B_{\lr i-1 \rr}| B_{\lr j \rr }, j > i-1)  = H(A_{\lr i-1 \rr}| A_{\lr j \rr}, j > i-1)   > 0.
$$ 

If  $\lr \alpha \rr \in \Omega$ such that 
$
H(B_{\lr \alpha \rr}| B_{\lr j \rr }, j > i-1)   > 0
$, then  
\[
H(B_{\lr \alpha \rr})  = H(A_{\lr \alpha \rr}) \ge H(A_{\lr i-1 \rr}) = H(B_{\lr i-1 \rr}).
\] 

Recall that 
$B_{\lr j \rr} = A^{*}_{\lr j \rr}$ for all $j > i -1$. 
By invoking Proposition \ref{prop1}, we  show that 
$B_{\lr i-1 \rr} = A^{*}_{\lr \omega_{\lr i-1 \rr} \rr}$ is an indicator variable for an element 
$\ell \in \X^{*} $ such that 
\[
\Pr(X^{*} = \ell ) = q_{i-1}.
\] 
We thus prove the induction step. As such, by properly renaming the elements, we can assume without loss of generality that 
$B_{\lr i-1 \rr} = A^{*}_{\lr i-1\rr}$ is an indicator variable for element $i-1$. 
So far, we have proved that $B_{\lr i \rr} = A^{*}_{\lr i \rr}$
for $i\ge 2$.

Next, we will also prove that $B_{\lr 1 \rr}$ is also an indicator binary random variable.
Recall from Proposition \ref{prop1} that   
\[
H(A_{\lr 1 \rr} | A_{\lr j \rr}, j \in \beta ) > 0
\]
for all proper subset $\beta$ of $[2,n]$. 
Hence, 
\[
H(B_{\lr 1 \rr} | A^{*}_{\lr j \rr}, j \in \beta ) > 0
\]
for all proper subset $\beta$ of $[2,n]$.
Invoking Proposition \ref{prop1}, then we can conclude that 
$B_{\lr 1 \rr}$ is the indicator random variable $A^{*}_{\lr 1 \rr}$.

Finally, we prove the equivalence property. 
Previously, we prove that  $B_{\lr 1 \rr}, B_{\lr 2 \rr}, \ldots, B_{\lr n\rr}$ are all distinct indicator random variables. Furthermore,    $q_{k}$ is the unique value between [0,1/2] such that 
\[
h_{b}( q_{k}) = H(B_{\lr k \rr})
\]
for all $k \ge 2$. As $H(B_{\lr k \rr}) = H(A_{\lr k \rr})$, we prove that 
$q_{k} = p_{k}$. Therefore, the distribution of $X$ and $X^{*}$ are essentially the same.

\section{Proof of Theorem 4 - Random Vector Case}\label{app.C}

\def\X{{\cal X}}

\def\bx{{\bf x}}
\def\by{{\bf y}}
\newcommand{\supp}[1]{{\mathcal S}(#1)}
\newcommand\msim[1]{{\: \sim_{#1} \:}}

In this appendix, we will prove Theorem \ref{thm2}, which extends Theorem \ref{thm1} to the random vector case. 

Consider a random vector  
\begin{align*}
X = (X_m : m \in \N_{M}).
\end{align*}
We will only consider the general case where its support is $\X$.

If $|\X|=1$, then all $H(X_m : m \in \N_{M}) = 0$. The theorem holds immediately.
If $|\X|=2$, then $H(X_{i})$ is either equal to $0$ or  $H(X_m : m \in \N_{M}) = 0$. If we let $\alpha$ be the subset of $\N_{M}$ such that 
$H(X_{i}) > 0 $ if and only if $i\in\alpha$. Then 
$X_{j}$ is deterministic (i.e., has zero entropies for all $j\not \in \alpha$). In addition, for any $i, j \in \alpha$, 
$X_{i}$ and $X_{j}$ are essentially the same (up to relabelling), or more precisely
\begin{align*}
H(X_{i} |X_{j}) = H(X_{j} |X_{i}) = 0.
\end{align*}
As each $X_{i}$ is binary, its distribution is also precisely characterised by the entropies.  Hence, the joint probability distribution of $(X_m : m \in \N_{M})$ is also well characterised. 

In the remaining of this appendix, we will assume that  
the size of $\X$  is at least 3, i.e.,  
$
{\cal S}(X_m : m \in \N_{M}) \ge 3
$.
Let $\X$ be the supports of $X$. 
Hence, elements of $\X$ is of the form 
$x=(x_{1},\ldots, x_{M})$ such that 
$$
\Pr(X_{m} = x_{m} , m \in \N_{M}) > 0
$$
if and only if $x \in \X$.


%
The collection of binary partition random variables induced by the random vector
$X = (X_{m}, m\in \N_{M})$ is again indexed by 
$
(A_{\lr \alpha \rr}, \lr \alpha \rr \in \Omega).
$
As before, we may assume without loss of generality  that  
\begin{equation*} 
A_{\lr \alpha \rr} = \left\{
  \begin{array}{l l}
    \alpha & \quad \text{if $X \in \alpha$}\\
    \alpha^{c} & \quad \text{otherwise.}\\
  \end{array} \right.
\end{equation*}

Now, suppose  
$$
(B_{\lr \alpha \rr} , \lr \alpha \rr \in \Omega)
$$ 
is a set of random variables satisfying the properties as specified in Theorem \ref{thm2}.
Invoking Theorem \ref{thm1} (by treating the random vector $X^{*}$ as one discrete variable), we can prove the following
\begin{enumerate}
\item
The size of the support of $X^{*}$ and  $X$ are the same.

\item
$
B_{\lr \alpha \rr} 
$ 
is a binary partition random variable for all $\lr\alpha\rr \in \Omega$.

\item 
The set of variables $(B_{\lr \alpha \rr} , \lr\alpha\rr \in \Omega)$ contains all distinct binary partition random variables induced by $X^{*}$.

\item $B_{\lr x \rr}$ is an indicator variable for all $x\in \X$. 

\end{enumerate}

Let $\X^{*}$ be the support of $X^{*}$. We similarly define 
$\Omega^{*}$ as the collection of all sets of the form $\{ \gamma , \gamma^{c} \}$ where 
$\gamma$ is a subset of $\X^{*}$ and the sizes of $\gamma$ and $\gamma^{c}$ are non-zero. Again, we will use $\lr \gamma \rr$ to denote the set and  define  
\begin{equation*} 
A^{*}_{\lr \gamma \rr} = \left\{
  \begin{array}{l l}
    \gamma & \quad \text{if $X^{*} \in \gamma$}\\
    \gamma^{c} & \quad \text{otherwise.}\\
  \end{array} \right.
\end{equation*}
According to definition,  $A_{\lr x \rr}$  is defined as an indicator variable for detecting $x$. However, while $B_{\lr x \rr}$ is an indicator variable, the subscript $x$ in $B_{\lr x \rr}$ is only an index. The element detected by $B_{\lr x \rr}$ can be any element in the support  of  $X^{*}$, which can be completely different from  $\X$. 
More precisely, we prove only the existence of  a mapping 
\[
\sigma : \X \mapsto \X^{*} 
\]
such that $B_{\lr x \rr}$ is an indicator random variable for 
detecting $\sigma(x)$. In other words
\begin{align*}
 B_{\lr x \rr} = A^{*}_{\lr \sigma(x) \rr}.
\end{align*}
In addition, for any $\lr\alpha\rr  \in \Omega$,  $B_{\lr \alpha \rr}$ is a binary partition random variable of $X^{*}$. 
For notation simplicity, we extend\footnote{
Strictly speaking, $\sigma(\alpha)$ is not precisely defined. As $\lr \gamma \rr = \lr \gamma^{c} \rr$, $\sigma(\alpha)$ can either be  $\gamma$ or $\gamma^{c}$. Yet, the precise choice of $\sigma(\alpha)$ does not have any effects on the proof. However, we only require that when $\alpha$ is a singleton, $\sigma(\alpha)$ should also be a singleton.  
} the mapping $\sigma$ such that 
\begin{align*}
A^{*}_{\lr \sigma(\alpha) \rr} = B_{\lr \alpha \rr}
\end{align*}
for all $ \alpha  \subseteq \X$.

\begin{remark}
When $n$ (the size of $\X$ and hence also the size of $\X^{*}$) is at least 3, there are exactly $n$ indicator random variables. As we shall see, we can use the indicator random variables to ``represent'' elements of $\X$ and $\X^{*}$ such that  their entropies  will determine the probability mass of each element in $\X$ and $\X^{*}$. 
\end{remark}

The lemma below follows from  Theorem \ref{thm1}. 
\begin{lemma}
For all $x\in \X$,  
$$
\Pr(X=x) = \Pr(X^{*} = \sigma(x))
$$
\end{lemma}

\begin{proposition}\label{prop2}
Let $\lr \alpha \rr \in \Omega$. Suppose  
$A_{\lr\beta\rr}$ satisfies the following properties:  
\begin{enumerate}
\item For any  $\gamma \subseteq \alpha$, 
$H(A_{\lr \beta \rr} |  A_{\lr x \rr} , x \in \gamma ) = 0$
if and only if $\gamma = \alpha$

\item For any  $\gamma \subseteq \alpha^{c}$, 
$H(A_{\lr \beta \rr} |  A_{\lr x \rr} , x \in \gamma ) = 0$
if and only if $\gamma = \alpha^{c}$.

\end{enumerate}
Then $A_{\lr \beta \rr} = A_{\lr \alpha \rr}$.
\end{proposition}
\begin{IEEEproof}
Direct verification.
\end{IEEEproof}

By construction of $B_{\lr \alpha \rr}$ (see \eqref{thm5:eqa}) and Proposition \ref{prop2}, we have the following result.
\begin{proposition}\label{prop3}
Let $\lr \alpha \rr \in \Omega$. Then  
$B_{\lr \beta \rr} = B_{\lr \alpha \rr}$ is the only binary partition variable of $X^{*}$ such that  
\begin{enumerate}
\item For any  $\gamma \subseteq \alpha$, 
$H(B_{\lr \beta \rr} |  B_{\lr x \rr} , x \in \gamma ) = 0$
if and only if $\gamma = \alpha$

\item For any  $\gamma \subseteq \alpha^{c}$, 
$H(B_{\lr \beta \rr} |  B_{\lr x \rr} , x \in \gamma ) = 0$
if and only if $\gamma = \alpha^{c}$.

\end{enumerate}
\end{proposition}

In the following proposition, we  further obtain properties about the mapping 
$\sigma$.

\begin{proposition}\label{prop4}
Let  $\alpha \in \X$ and 
$
\delta(\alpha) = \{ \sigma(x) : x\in\alpha  \}
$. 
Then  
\[
\lr \sigma(\alpha) \rr = \lr \delta(\alpha) \rr.
\]
\end{proposition}
\begin{IEEEproof}
By Proposition \ref{prop3},   $B_{\lr \alpha \rr} = A^{*}_{\lr \sigma(\alpha) \rr}$    
  is the only variable such that 
\begin{enumerate}
\item For any  $\gamma \subseteq \alpha$, 
$H(A^{*}_{\lr \sigma(\alpha) \rr} |  A^{*}_{\lr \sigma(x) \rr} , x \in \gamma ) = 0$
if and only if $\gamma = \alpha$

\item For any  $\gamma \subseteq \alpha^{c}$, 
$H(A^{*}_{\lr \sigma(\alpha) \rr} |  A^{*}_{\lr \sigma(x) \rr} , x \in \gamma ) = 0$
if and only if $\gamma = \alpha^{c}$.
\end{enumerate}

The above two properties can then be rephrased as 
\begin{enumerate}
\item For any  $\delta(\gamma) \subseteq \delta(\alpha)$, 
$$
H(A^{*}_{\lr \sigma(\alpha) \rr} |  A^{*}_{\lr \sigma(x) \rr} ,  \sigma(x) \in \delta(\gamma) ) = 0
$$
if and only if $\delta(\gamma) = \delta(\alpha)$

\item For any  $\delta(\gamma) \subseteq  \delta(\alpha^{c})$, 
$$
H(A^{*}_{\lr \sigma(\alpha) \rr} |  A^{*}_{\lr \sigma(x) \rr} , \sigma(x) \in \delta(\gamma) ) = 0
$$
if and only if $\delta(\gamma) = \delta(\alpha^{c})$.
\end{enumerate}
Now, we can invoke Proposition \ref{prop2} again and prove that  
\[
A^{*}_{\lr \delta(\alpha) \rr} = A^{*}_{\lr \sigma(\alpha) \rr}
\]
or equivalently, 
$
\lr\delta(\alpha) \rr = \lr \sigma(\alpha) \rr
$.
The proposition then follows.
\end{IEEEproof}

\begin{remark}
Due to Proposition \ref{prop4}, we will assume in the remaining of the paper that 
\begin{align*}
\sigma(\alpha) = \delta(\alpha) =\{ \sigma(x) : \: x\in\alpha \}.
\end{align*}
\end{remark}
 
So far, we have proved very interesting properties about the 
mapping $\sigma$. In particular, we showed that for any $\alpha \subseteq \X$, $B_{\lr \alpha \rr} = A^{*}_{\lr \delta(\alpha) \rr}$ 
where $\delta(\alpha) = \{ \alpha(x) :\: x\in\alpha \}$. Hence, we now know all the entropies of the binary partition random variables $A^{*}_{\lr \delta(\alpha) \rr}$. Furthermore, by the construction of $B_{\lr \alpha \rr}$, we have 
\begin{align*}
A_{\lr \alpha \rr} = A^{*}_{\lr \delta(\alpha) \rr}. 
\end{align*}
In other words, the entropies of the binary partition random variables 
$A_{\lr \alpha \rr}$ and $A^{*}_{\lr \delta(\alpha) \rr}$ are the same.
In the following, we will show that the joint probability distributions of the two random vectors $X$ and $X^{*}$ are the same (up to relabelling).

\begin{proposition}\label{prop5}
Consider two distinct elements $x=(x_{1},\ldots, x_{M})$ and 
$x'=(x'_{1},\ldots, x'_{M})$ in $\X$. 
Let 
\begin{align}
\sigma(x) & =  y = (y_{1}, \ldots, y_{M})  \label{eq:87}\\
\sigma(x') & = y' = (y'_{1}, \ldots, y'_{M}). \label{eq:88} 
\end{align}
Then 
$
x_{m} \neq x'_{m}
$
if and only if 
$
y_{m} \neq y'_{m}
$.
\end{proposition}
\begin{IEEEproof}
First, we will prove the only-if statement. 
Suppose $x_{m} \neq x'_{m}$. 
 Consider the following two sets
 \begin{align*}
 \Theta = \{x''=(x''_{1},\ldots, x''_{M}) \in \X : \: x''_{m} \neq x_{m} \}
 \end{align*}
 and 
 \begin{align*}
 \Theta^{c} = \{x''=(x''_{1},\ldots, x''_{M}) \in \X  : \: x''_{m} = x_{m} \}.
 \end{align*}
It is obvious that 
$
H(A_{\lr\Theta\rr} | X_{m}) = 0.
$
By \eqref{eq34}-\eqref{eq35}, we have  
$
H(B_{\lr\Theta\rr} | X^{*}_{m}) = 0
$.
By  definition, we prove that 
$
B_{\lr\Theta\rr} = A^{*}_{\lr\sigma(\Theta) \rr}.
$
Since  
$
H(B_{\lr \Theta \rr} | X^{*}_{m}) = 0
$, 
this implies 
$
H( A^{*}_{\lr\sigma(\Theta) \rr} | X^{*}_{m}) = 0
$. 

Now, notice that $x \in \Theta^{c}$ and $x' \in \Theta$.
By Proposition \ref{prop4},  
$
\sigma(\Theta)  =  \{  \sigma(x) : x \in \Theta \}
$. 
 Therefore, 
$
y'=\sigma(x') \in \sigma(\Theta)
$
 and 
$
y=\sigma(x) \not\in \sigma(\Theta).
$
Together with the fact that 
$
H( A^{*}_{\lr\sigma(\Theta\nu) \rr} | X^{*}_{m}) = 0
$, 
we can then prove that 
$$
y'_{m} \neq y''_{m}.
$$

Next, we prove the if-statement. Suppose $y, y' \in \X^{*}$ such that  
$
y_{m} \neq y'_{m}
$. 
There exist  $x$ and $x'$ such that \eqref{eq:87} and \eqref{eq:88} hold. 
Again, define 
 \begin{align*}
 \Lambda = \{y''=(y''_{1},\ldots, y''_{M}) \in \X^{*} : \: y''_{m} \neq y_{m} \}
 \end{align*}
 and 
 \begin{align*}
 \Lambda^{c} = \{y''=(y''_{1},\ldots, y''_{M}) \in \X^{*}  : \: y''_{m} = y_{m} \}.
 \end{align*}
Then 
$
H(A^{*}_{\lr \Lambda \rr} | X^{*}_{m}) = 0
$.
Let 
\[
\Xi = \{ x\in \X :\: \sigma(x) \in \Lambda\}.
\]
By definition and Proposition \ref{prop4} 
\begin{align*}
B_{\lr \Xi \rr} & = A^{*}_{\lr \sigma (\Xi)  \rr} \\
& = A^{*}_{\lr  \Lambda \rr} .
\end{align*}
Hence, we have $H(B_{\lr\Xi\rr} | X^{*}_{m}) = 0$ and consequently 
$H(A_{\lr\Xi\rr} | X_{m}) = 0$. 

On the other hand, it can be verified from definition that  $x \in \Xi^{c}$ and $x' \in \Xi$. Together with that  
$H(A_{\lr\Xi\rr} | X_{m}) = 0$, we prove that   $x_{m} \neq x'_{m}$. The proposition then follows.
\end{IEEEproof}

We have now proved all the necessary intermediate results. The proof for Theorem  \ref{thm2} is given as follows.
 
\begin{IEEEproof}[Proof of Theorem \ref{thm2}]
A direct consequence of Proposition \ref{prop5} is that there exists bijective mappings $\sigma_{1} , \ldots, \sigma_{M}$ such that 
\begin{align*}
\sigma(x) = (\sigma_{1}(x_{1}), \ldots, \sigma_{M}(x_{M})).
\end{align*} 
On the other hand, Theorem \ref{thm1} proved that 
$$
\Pr( X = x) = \Pr ( X^{*} = \sigma(x)).
$$
Consequently, 
$$\Pr(X_{1} = x_{1}, \ldots,  X_{M} = x_{M}) = \Pr(X^{*}_{1} = \sigma_{1}(x_{1}), \ldots,  X^{*}_{M} = \sigma_{M}(x_{M})). $$
Therefore,  the joint distributions of $X= (X_{1} , \ldots, X_{M})$ and 
$X^{*}= (X^{*}_{1} , \ldots, X^{*}_{M})$ are essentially the same (by renaming $x_{m}$ as $\sigma_{m}(x_{m})$).
\end{IEEEproof}

\end{appendices}

\bibliographystyle{ieeetr}
\bibliography{network}

\begin{thebibliography}{10}

\bibitem{ThaChaGra11}
S.~Thakor, T.~Chan, and A.~Grant, ``Bounds for network information flow with
  correlated sources,'' in {\em Australian Communications Theory Workshop
  (AusCTW)}, (Melbourne, Australia), pp.~43 --48, Feb. 2011.

\bibitem{ThaChaGra13}
S.~Thakor, T.~Chan, and A.~Grant, ``Characterising correlation via entropy
  functions,'' in {\em Information Theory Workshop (ITW), 2013 IEEE}, pp.~1--2,
  Sept 2013.

\bibitem{ThaChaGra16}
S.~Thakor, T.~Chan, and A.~Grant, ``Characterising probability distributions
  via entropies,'' in {\em International Symposium on Information Theory and
  its Applications (accepted)}, (California, USA), Oct-Nov. 2016.

\bibitem{ChaGra08}
T.~H. Chan and A.~Grant, ``Dualities between entropy functions and network
  codes,'' {\em IEEE Trans. Inform. Theory}, vol.~54, pp.~4470--4487, Oct.
  2008.

\bibitem{AhlCai00}
R.~Ahlswede, N.~Cai, S.-Y.~R. Li, and R.~W. Yeung, ``Network information
  flow,'' {\em IEEE Trans. Inform. Theory}, vol.~46, pp.~1204--1216, July 2000.

\bibitem{YeuZha99}
R.~W. Yeung and Z.~Zhang, ``On symmetrical multilevel diversity coding,'' {\em
  IEEE Trans. Inform. Theory}, vol.~45, pp.~609--621, Mar. 1999.

\bibitem{Yeu08}
R.~W. Yeung, {\em Information Theory and Network Coding}.
\newblock Springer, 2008.

\bibitem{LiYeu03}
S.-Y.~R. Li, R.~Yeung, and N.~Cai, ``Linear network coding,'' {\em IEEE Trans.
  Inform. Theory}, vol.~49, pp.~371--381, Feb. 2003.

\bibitem{SleWol73}
D.~Slepian and J.~Wolf, ``Noiseless coding of correlated information sources,''
  {\em IEEE Trans. Inform. Theory}, vol.~19, pp.~471--480, Jul 1973.

\bibitem{WynZiv76}
A.~Wyner and J.~Ziv, ``The rate-distortion function for source coding with side
  information at the decoder,'' {\em IEEE Trans. Inform. Theory}, vol.~22,
  pp.~1--10, Jan 1976.

\bibitem{RamJaiChoEff06}
A.~Ramamoorthy, K.~Jain, P.~Chou, and M.~Effros, ``Separating distributed
  source coding from network coding,'' {\em IEEE Trans. Inform. Theory},
  vol.~52, pp.~2785 -- 2795, Jun. 2006.

\bibitem{Han11}
T.~S. Han, ``Multicasting multiple correlated sources to multiple sinks over a
  noisy channel network,'' {\em IEEE Trans. Inform. Theory}, vol.~57, pp.~4
  --13, Jan. 2011.

\bibitem{Han80}
T.~S. Han, ``Slepian-{W}olf-{C}over theorem for a network of channels,'' {\em
  Inform. Control}, vol.~47, no.~1, pp.~67--83, 1980.

\bibitem{BarSer06}
J.~Barros and S.~Servetto, ``Network information flow with correlated
  sources,'' {\em IEEE Trans. Inform. Theory}, vol.~52, pp.~155 -- 170, Jan.
  2006.

\bibitem{ThaGraCha16a}
S.~Thakor, A.~Grant, and T.~Chan, ``Cut-set bounds on network information
  flow,'' {\em IEEE Trans. Inform. Theory}, vol.~62, pp.~1850--1865, April
  2016.

\bibitem{ThaGraCha09}
S.~Thakor, A.~Grant, and T.~Chan, ``Network coding capacity: A functional
  dependence bound,'' in {\em IEEE Int. Symp. Inform. Theory}, pp.~263 --267,
  Jul. 2009.

\bibitem{CovTho06}
T.~M. Cover and J.~A. Thomas, {\em Elements of Information Theory}.
\newblock Wiley-Interscience, 2006.

\bibitem{Yeu02}
R.~W. Yeung, {\em A First Course in Information Theory}.
\newblock New York: Kluwer Academic/Plenum Publishers, 2002.

\bibitem{GohYanJagg13}
A.~Gohari, S.~Yang, and S.~Jaggi, ``Beyond the cut-set bound: Uncertainty
  computations in network coding with correlated sources,'' {\em IEEE Trans.
  Inform. Theory}, vol.~59, pp.~5708--5722, Sept 2013.

\bibitem{GohYanJagg11}
A.~Gohari, S.~Yang, and S.~Jaggi, ``Beyond the cut-set bound: Uncertainty
  computations in network coding with correlated sources,'' in {\em IEEE Int.
  Symp. Inform. Theory}, pp.~598--602, July 2011.

\bibitem{GacKor73}
P.~G\'{a}cs and J.~Korner, ``{Common information is far less than mutual
  information},'' {\em Probl. Inform. Control}, vol.~2, no.~2, pp.~149--162,
  1973.

\bibitem{Ren60}
A.~Renyi, ``{On measures of information and entropy},'' in {\em Proceedings of
  the 4th Berkeley Symposium on Mathematics, Statistics and Probability},
  pp.~547--561, 1960.

\bibitem{Tsa88}
C.~Tsallis, ``Possible generalization of boltzmann-gibbs statistics,'' {\em
  Journal of Statistical Physics}, vol.~52, no.~1, pp.~479--487, 1988.

\end{thebibliography}
\end{document}